\documentclass[preprint,12pt]{elsarticle}

\usepackage{subcaption}   

\usepackage{amssymb}
\usepackage{amsmath}
\usepackage{tabularx} 
\usepackage[section]{placeins}
\usepackage{caption}
\usepackage{graphicx}
\usepackage{booktabs}
\usepackage{siunitx} 
\usepackage{makecell}
\usepackage{bm}
\usepackage{amsmath}

\usepackage[hidelinks]{hyperref}

\begin{document}

\begin{frontmatter}



\title{Posterior-Guided Adaptive Surrogate Refinement for Efficient Bayesian Calibration of Transient Nuclear Energy Systems}


\author[uw]{Zikang Yao\corref{cor1}}
\author[uw]{Eugene Yee}
\author[uw]{Fue-Sang Lien}
\cortext[cor1]{Corresponding author. Email: z24yao@uwaterloo.ca}

\affiliation[uw]{
    organization={Department of Mechanical and Mechatronics Engineering, University of Waterloo},
    addressline={200 University Avenue West},
    city={Waterloo},
    postcode={N2L 3G1},
    state={Ontario},
    country={Canada}
}

\begin{abstract}
Credible uncertainty-informed simulation is essential for model validation, safety assessment, and digitalized operation of advanced nuclear energy systems. However, Bayesian calibration of full-transient thermal--hydraulic models remains computationally prohibitive because posterior inference requires repeated evaluations of expensive high-fidelity simulators. Although surrogate models can accelerate posterior sampling, conventional fixed surrogates may provide insufficient accuracy in posterior-relevant regions and consequently distort inferred parameter distributions and calibrated predictions. This study develops a posterior-guided adaptive inverse uncertainty quantification framework designed to preserve posterior fidelity under a limited high-fidelity simulation budget. The framework integrates principal component analysis for compression of multi-channel transient responses, Gaussian process regression for probabilistic surrogate modeling, Microcanonical Langevin Monte Carlo for posterior sampling, and a persistent extensible Sobol candidate reservoir. At each adaptive iteration, candidates are classified according to the geometry of the evolving posterior distribution and selected using GPR predictive uncertainty, distance from the existing training design, and within-batch diversity. A local--global strategy concentrates new simulations in posterior-relevant regions while retaining exploration of the original parameter domain. The framework is demonstrated on the NACIE system-scale natural-circulation benchmark using transient temperature and mass-flow-rate observations. Using 288 Flownex SE evaluations, the adaptive surrogate reproduces the principal characteristics of the direct Flownex-based posterior, including posterior means, uncertainty levels, marginal distributions, nonlinear parameter dependence, and posterior predictive responses. Relative to a fixed PCA--GPR surrogate trained using 1024 simulations, the proposed framework reduces the required high-fidelity training evaluations by 71.9\% while retaining comparable posterior accuracy. Relative to a direct reference requiring more than 10000 simulator evaluations, it reduces the high-fidelity model-evaluation cost by more than 97.1\%, corresponding to an approximately 34.7-fold reduction. System-level posterior analysis further reveals a hydraulic response direction most strongly associated with the resistance coefficient and a coupled thermal compensation direction involving ambient heat loss and heat-exchanger conductance. These results demonstrate that posterior-guided refinement can preserve both statistical and physically meaningful joint posterior structures while substantially reducing the cost of Bayesian calibration, providing an enabling methodology for repeated uncertainty-aware model updating and subsequent reliability and safety assessment.
\end{abstract}


\end{frontmatter}
\section{Introduction}

Advanced nuclear energy systems are expected to play an important role in future low-carbon energy portfolios because of their high energy density, stable power output, and potential for long-duration, dispatchable electricity generation. For these systems, reliable transient simulation is essential not only for safety analysis, but also for model validation, design-margin evaluation, system optimization, operational decision support, and uncertainty-informed assessment. In particular, system-scale thermal-hydraulic responses, including temperature evolution, flow redistribution, natural-circulation behavior, and heat-removal capability, are strongly affected by uncertain model parameters, boundary conditions, measurement errors, and model-form assumptions. If these uncertainties are not properly quantified, deterministic simulations may provide misleading confidence in predicted safety margins and system performance. Therefore, efficient calibration and uncertainty quantification of transient thermal-hydraulic models against experimental observations are key requirements for the credible assessment of advanced nuclear energy systems.

Meanwhile, nuclear energy research is undergoing a broader transition toward digitalized monitoring, intelligent prediction, and continuously updated computational models. Digital twins have increasingly been recognized as an important technology for integrating physical models, operational data, monitoring information, and predictive analysis throughout the reactor life cycle \cite{HUANG2025126922}. Recent studies have developed digital-twin frameworks for real-time operation monitoring and future-state prediction of space nuclear power systems \cite{ZHU2026111843}, as well as self-calibrating digital twins capable of adapting reactor models under fault conditions \cite{ZHANG2026111995}. These developments indicate that future nuclear digital twins must not only reproduce nominal system behavior, but also assimilate observations, update uncertain model characteristics, and provide credible predictive uncertainty as operating conditions evolve.

Substantial progress has also been made in data-driven nuclear-system monitoring and prediction. Pre-trained large language models have been applied to the forecasting of nuclear power plant operating parameters \cite{CHANG2026127438}, while integrated neural-network models have been developed for coupled multivariate time-series prediction and nuclear power plant accident prevention \cite{XIAO2024132374}. Uncertainty-aware deep-learning approaches have further been proposed for reliable health monitoring in safety-critical energy systems \cite{YAO2024130419}. In addition, integrated neural-network and optimization frameworks have recently been investigated for uncertainty-informed calibration of nuclear reactor thermal-hydraulic models \cite{XIONG2026112281}. These studies demonstrate the potential of modern data-driven methods for extracting predictive information from complex monitoring and simulation datasets.

Data-driven forecasting and physics-based model calibration, however, address different but complementary requirements. Data-driven models are primarily designed to learn predictive relationships from available observations, whereas physics-based inverse uncertainty quantification seeks to infer uncertain physical parameters of an established simulator and propagate the remaining uncertainty to unobserved or safety-relevant quantities of interest. Retaining the physical simulator is particularly valuable when parameter interpretability, conservation-law consistency, extrapolation beyond observed operating conditions, or accident-transient prediction is required. A key methodological challenge is therefore to combine the computational efficiency of data-driven surrogate models with the interpretability and uncertainty-accounting capabilities of physics-based system models.

This challenge is directly relevant to reliability and nuclear safety assessment. Recent studies have evaluated accident-tolerant fuel performance under loss-of-coolant accident conditions \cite{XIONG2024109847,KONG2026141113}, investigated the dynamic performance of coordinated control systems for small pressurized water reactors using uncertainty quantification and sensitivity analysis \cite{CHEN2025110794}, and developed efficient strategies for probabilistic safety-margin analysis under small-break loss-of-coolant accidents \cite{CHEN2026141470}. These applications demonstrate that uncertainties in thermal-hydraulic parameters, physical models, and operating conditions can materially affect predicted accident responses, control-system performance, and probabilistic safety margins. Credible estimation of such quantities requires not only accurate nominal simulations, but also a statistically consistent characterization of parameter uncertainty and parameter dependence.

Bayesian inverse uncertainty quantification (IUQ) provides a rigorous framework for inferring uncertain model parameters from experimental data and propagating the resulting uncertainty to model predictions. Following the Bayesian calibration formulation of Kennedy and O'Hagan~\cite{ref1}, IUQ has been extended to complex computer models with high-dimensional, functional, and spatio-temporal outputs through basis representations, reduced-order expansions, and dimension-reduction techniques~\cite{ref2}. Rather than identifying only a single optimal parameter vector, Bayesian IUQ estimates the joint posterior distribution of uncertain parameters after accounting for prior information, experimental observations, measurement uncertainty, and residual model discrepancy. The resulting posterior distribution characterizes both individual parameter uncertainties and nonlinear dependence among parameters, and can subsequently be propagated to obtain calibrated predictive distributions.

This property makes Bayesian IUQ particularly attractive for transient energy-system simulations, where the quantities of interest are not isolated scalar responses but coupled time-dependent signals measured at multiple locations. However, direct Bayesian calibration remains computationally prohibitive for high-fidelity thermal-hydraulic models. A single full-transient simulation may require substantial computational time \cite{YAO2026140531}, while posterior exploration generally requires thousands to millions of likelihood evaluations. As a result, direct high-fidelity Bayesian IUQ is rarely practical for routine uncertainty-informed analysis, repeated model updating, and digital assessment of system-scale nuclear energy models~\cite{ref15}.

Surrogate modeling has become a widely used strategy to alleviate this computational burden. Gaussian process regression (GPR), polynomial chaos expansion, reduced-order modeling, and machine-learning-based emulators have been used to replace expensive simulations during posterior sampling~\cite{ref9,ref21}. In nuclear thermal-hydraulic applications, surrogate-assisted Bayesian IUQ has been successfully applied to model calibration problems, including system-code calibration and TRACE-related uncertainty quantification studies~\cite{ref14,ref10}. These approaches can substantially reduce the cost of posterior sampling once an accurate surrogate model is available.

Nevertheless, the construction of such a surrogate introduces another computational bottleneck. A fixed surrogate is usually trained using samples generated before Bayesian inference begins, and its posterior accuracy depends strongly on whether the training data adequately cover the posterior-relevant region. If the initial design is insufficient in regions of high posterior probability, the surrogate may introduce biased posterior means, inaccurate uncertainty bounds, or distorted parameter correlations. Such errors can be particularly important in reliability-oriented applications because an apparently accurate surrogate according to global validation metrics may still be inaccurate in the limited parameter region that dominates the likelihood and posterior distribution.

This limitation is especially important for complex transient thermal-fluid energy systems, including advanced nuclear energy systems, natural-circulation loops, integrated energy systems, and other multi-physics applications. The mapping from uncertain input parameters to transient responses is often nonlinear, multi-output, and strongly time-dependent. The evolution of one quantity, such as the system mass flow rate, may be tightly coupled with temperature distributions, heat-removal capability, and energy-transport processes in other parts of the system. Consequently, constructing surrogate models for such problems usually requires both output dimension reduction and a sufficiently large number of high-fidelity simulations~\cite{ref25}.

Principal component analysis combined with Gaussian process regression provides an effective representation for such high-dimensional transient outputs. PCA compresses the original multi-channel time-series responses into a small number of latent coefficients, while separate GPR models map the uncertain physical parameters to the retained coefficients. The complete transient responses can then be reconstructed during Bayesian inference. In addition to efficient prediction, GPR provides an estimate of epistemic predictive uncertainty, which can be used to identify parameter regions where the current surrogate remains insufficiently informed.

Although increasing the training dataset can improve global surrogate accuracy, every additional sample still requires an expensive system-level simulation. Therefore, the computational bottleneck is gradually shifted from posterior sampling to high-fidelity data generation. This creates a fundamental efficiency challenge for surrogate-assisted IUQ: surrogate models are needed to make Bayesian calibration computationally affordable, while the construction of a uniformly accurate surrogate may itself demand a prohibitively large simulation budget.

More importantly, minimizing the average surrogate error over the complete prior domain is not necessarily the most appropriate objective for Bayesian IUQ. Surrogate errors in regions with negligible posterior probability may have little influence on the final calibration result. In contrast, relatively small errors in regions that strongly affect the likelihood may alter posterior means, posterior variances, parameter correlations, and posterior predictive intervals. The primary objective of surrogate-assisted IUQ is therefore not simply global emulator accuracy, but preservation of the inferred posterior distribution and the associated calibrated transient predictions.

Existing active-learning and adaptive-sampling methods have been developed to reduce the cost of surrogate construction in engineering applications~\cite{ref28}. However, many of these strategies are primarily designed to improve surrogate accuracy globally over the input domain, reduce prediction error in an average sense, or identify conventional limit-state boundaries. For Bayesian IUQ of complex transient system simulators, these objectives are not necessarily optimal. Surrogate fidelity is most valuable in regions that strongly influence the likelihood and posterior probability. Refining the surrogate uniformly over the entire prior space may consume expensive high-fidelity simulations in regions that have little impact on the calibrated model. A more efficient strategy is to place new simulations where the surrogate uncertainty is large and where surrogate approximation errors are most likely to affect posterior inference.

Several additional challenges must also be considered. First, the posterior-relevant region is unknown before the initial Bayesian analysis and may change as the surrogate is refined. Second, aggressive concentration around an early approximate posterior may prevent the algorithm from discovering other plausible regions of the parameter space. Third, the value of a candidate point depends not only on surrogate predictive uncertainty, but also on its distance from existing training points and its redundancy with other points selected in the same adaptive batch. Finally, convergence should be assessed according to the stability of posterior statistics and calibrated transient responses, rather than global surrogate error alone.

To address these challenges, this study develops a posterior-guided adaptive IUQ framework for expensive transient energy-system simulations, with advanced nuclear thermal-hydraulic analysis used as a representative application. The proposed method couples two complementary sources of information during surrogate refinement. The first is the epistemic uncertainty provided by the GPR surrogate, which identifies regions where the current emulator remains unreliable. The second is the posterior information obtained from Bayesian sampling, which identifies parameter regions that are most influential for calibration. New high-fidelity simulations are preferentially selected in regions that are both uncertain for the surrogate and relevant to the posterior distribution. In this way, the surrogate is refined where approximation errors are most likely to alter the Bayesian calibration result, rather than being improved uniformly over the admissible parameter space.

The proposed framework integrates PCA-based transient-output compression, GPR surrogate modeling, Microcanonical Langevin Monte Carlo (MCLMC) posterior sampling, an extensible Sobol candidate design, and posterior-guided adaptive surrogate refinement. A single Sobol sequence is generated over the fixed prior domain. The initial subset is evaluated using the high-fidelity simulator, while the remaining unevaluated points form a persistent candidate reservoir. During each adaptive iteration, the retained candidates are reclassified and rescored according to the updated posterior and surrogate information rather than being discarded.

Posterior samples are used to characterize the scale, orientation, and correlation structure of the current posterior distribution. A Mahalanobis-distance criterion separates candidates located within the principal posterior-support region from candidates retained for global exploration. Candidate selection then combines GPR predictive uncertainty, distance from the existing high-fidelity training design, and within-batch diversity. The local component concentrates computational resources near the current posterior support, whereas the global component preserves exploration over the original admissible domain. This local--global strategy reduces the risk of premature concentration around an inaccurate early posterior and improves the robustness of the adaptive process.

The framework is demonstrated on the NACIE natural-circulation benchmark, which serves as a representative system-scale transient thermal-fluid problem relevant to advanced nuclear energy applications. Unlike simplified analytical examples or low-dimensional test functions, the NACIE case involves coupled heat-transfer and flow mechanisms, multi-channel time-dependent observations, nonlinear parameter interactions, and expensive high-fidelity simulations. These characteristics make it a demanding test case for assessing whether posterior-guided adaptive surrogate refinement can make Bayesian IUQ practical for realistic system models.

The proposed framework is validated against direct Flownex-based Bayesian IUQ and compared with a fixed PCA--GPR surrogate strategy. The assessment considers posterior means, posterior standard deviations, marginal distributions, uncertainty levels, nonlinear parameter correlations, posterior predictive responses, and computational cost. The results show that the adaptive method reproduces the principal characteristics of the direct high-fidelity posterior while requiring substantially fewer high-fidelity simulations. This demonstrates that posterior-guided adaptive surrogate refinement can reduce the data-generation cost of Bayesian calibration while preserving the probabilistic information required for credible model assessment.

The framework does not itself constitute a complete operational digital twin or probabilistic safety assessment platform. Instead, it addresses a central enabling requirement of such applications: efficient and uncertainty-aware updating of an expensive physics-based transient model. The resulting calibrated posterior can provide a more credible probabilistic basis for subsequent safety-margin evaluation, reliability analysis, digital model updating, and uncertainty-informed operational assessment.

The main contributions of this work are summarized as follows:

\begin{enumerate}
    \item A posterior-guided adaptive surrogate refinement framework is developed for Bayesian IUQ of expensive, safety-critical transient system models.

    \item GPR predictive uncertainty and posterior relevance are combined to select high-fidelity simulations in regions that are most influential for Bayesian calibration.

    \item A persistent extensible Sobol candidate design and a local--global classification strategy are introduced to balance posterior-focused refinement and global exploration without contracting the original prior domain.

    \item Training-design distance and within-batch diversity are incorporated to reduce redundant sampling and improve the information gained from each high-fidelity simulation.

    \item PCA--GPR surrogate modeling is integrated with MCLMC posterior sampling to efficiently handle high-dimensional, multi-output transient responses.

    \item The method is validated on the NACIE system-scale benchmark and shown to reproduce direct Flownex-based posterior inference while substantially reducing the required number of high-fidelity simulations.
\end{enumerate}

The proposed framework therefore provides a practical route for uncertainty-informed calibration, model validation, and credible digital assessment of advanced nuclear energy systems and other complex transient thermal-fluid applications. It also provides an enabling methodological foundation for repeated model updating, uncertainty-aware digital twins, and subsequent probabilistic safety and reliability assessment.

\section{Problem Definition and Description}

\subsection{NACIE Thermal-Hydraulic System}

The Natural Circulation Experiment (NACIE) facility\cite{coccoluto2011nacie} is a representative system-scale thermal-hydraulic benchmark developed for investigating natural circulation phenomena in heavy liquid metal reactor systems. The facility reproduces the dominant thermal-hydraulic mechanisms encountered in lead-cooled fast reactors, including buoyancy-driven circulation, heat transfer, hydraulic resistance, and thermal inertia effects.

\begin{figure}
    \centering
    \includegraphics[width=1\linewidth]{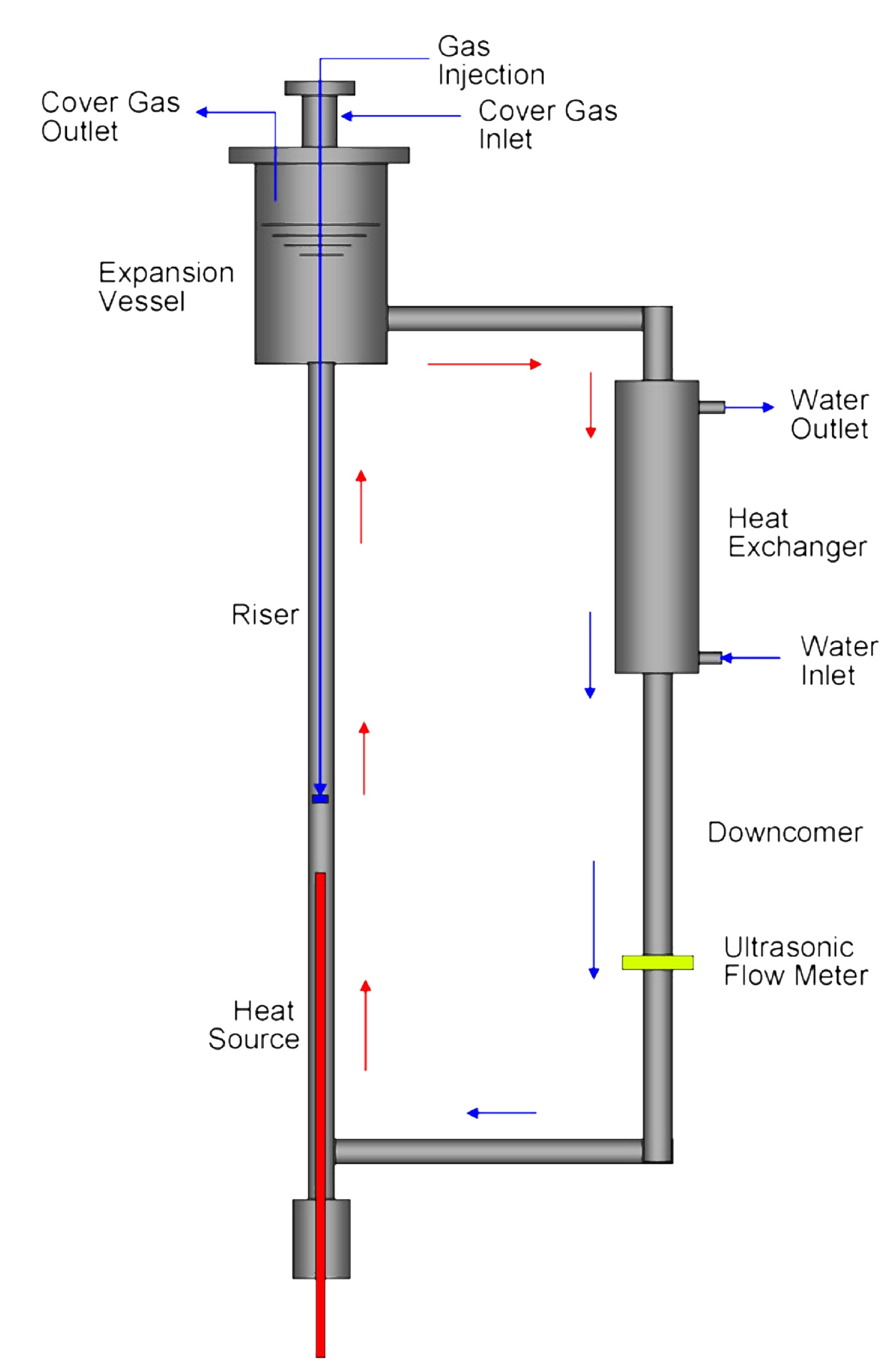}
    \caption{Layout of the NACIE loop experimental configuration\cite{coccoluto2011nacie}.}
    \label{fig:nacie layout}
\end{figure}

As shown in  Fig.~\ref{fig:nacie layout} the NACIE loop consists of a heater section, a heat exchanger, riser and downcomer branches, and connecting piping components. During operation, thermal energy supplied by the heater generates density differences within the lead-bismuth eutectic (LBE) coolant, thereby inducing natural circulation throughout the loop. The resulting flow and temperature evolution are governed by strongly coupled thermal-hydraulic processes involving momentum transport, heat transfer, and buoyancy feedback.
In this study, transient experimental measurements obtained from the NACIE facility are used for inverse uncertainty quantification. The quantities of interest include temperature responses at multiple measurement locations and the corresponding loop mass flow rate. Unlike conventional component-level or steady-state benchmarks, the NACIE facility provides a full-transient system-level thermal-hydraulic test case that captures the coupled evolution of flow and heat transfer phenomena throughout the entire loop. The resulting response is governed by nonlinear interactions among buoyancy-driven circulation, distributed thermal losses, hydraulic resistance, and system thermal inertia across multiple temporal and spatial scales. Consequently, NACIE serves as a representative benchmark for investigating Bayesian calibration and adaptive inverse uncertainty quantification in complex full-scale thermal-hydraulic systems.

\subsection{Problem Formulation}

The objective of this work is to infer uncertain thermal-hydraulic model parameters using experimental observations and Bayesian inverse uncertainty quantification.

The uncertain parameter vector is defined as

\begin{equation}
\theta
=
[P_b,C_k,h_a,\lambda_p]
\end{equation}
where

\begin{itemize}

\item $P_b$ denotes the heater power bias.

\item $C_k$ denotes the hydraulic resistance coefficient.

\item $h_a$ denotes the ambient heat transfer coefficient.

\item $\lambda_p$ denotes the thermal conductivity.

\end{itemize}

The experimental observations are represented by

\begin{equation}
D=
\left\{
y_c^{obs}(t_j)
\right\},
\qquad
c=1,\ldots,N_c,
\quad
j=1,\ldots,N_t.
\end{equation}
where $N_c$ and $N_t$ denote the number of observation channels and time step, respectively, and $t_j$ represents the $j$-th time step.

\begin{equation}
N_D=N_cN_t.
\end{equation}
The dataset $D$ represents the complete multi-output transient experimental observations used for Bayesian calibration.

For a given parameter vector $\theta$, the corresponding transient response predicted by the thermal-hydraulic model is

\begin{equation}
y
=
\eta(\theta)
\end{equation}
where $\eta(\theta)$ denotes the Flownex SE simulator and $y$ represents the complete simulated multi-output transient response vector.

The objective of inverse uncertainty quantification is to estimate the posterior distribution of uncertain parameters through Bayesian inference,

\begin{equation}
p(\theta|D)
\propto
p(D|\theta)p(\theta)
\end{equation}
where $p(\theta)$ is the prior distribution, $p(D|\theta)$ is the likelihood function, and $p(\theta|D)$ is the posterior distribution.

\subsection{Computational Challenges}

Although Bayesian inverse uncertainty quantification provides a rigorous framework for parameter calibration and uncertainty characterization, its application to system-scale transient thermal-hydraulic simulations remains computationally demanding.

For the NACIE system, each model evaluation requires solving the complete transient evolution of the thermal-hydraulic loop over the entire simulation period. Consequently, a single high-fidelity simulation may require several minutes of computational time.

The total computational cost associated with Bayesian calibration can be approximated as

\begin{equation}
C_{total}
=
N_{eval}
\times
C_{HF}
\end{equation}
where $N_{eval}$ denotes the number of likelihood evaluations and $C_{HF}$ denotes the computational cost of a single high-fidelity simulation.

For practical inverse uncertainty quantification problems, the required number of model evaluations may reach tens of thousands or even hundreds of thousands. When high-fidelity thermal-hydraulic simulators are employed, such repeated evaluations often become computationally prohibitive due to the significant cost associated with each transient simulation.

To alleviate this limitation, surrogate-assisted inverse uncertainty quantification frameworks have been widely developed. By replacing the original simulator with computationally efficient surrogate models, such as PCA-GPR, FPCA-GPR, neural networks, and other machine-learning-based approaches, the computational cost associated with posterior sampling can be reduced by several orders of magnitude. However, despite the substantial acceleration achieved during Bayesian inference, the construction of accurate surrogate models remains heavily dependent on high-fidelity simulation data. For complex system-scale transient problems such as NACIE and more realistic large-scale configurations such as NACIE-UP, generating the surrogate training dataset itself may require hundreds or even thousands of expensive thermal-hydraulic simulations. Consequently, the computational bottleneck is gradually shifted from Bayesian posterior exploration to high-fidelity data generation.

Furthermore, achieving higher surrogate accuracy generally requires larger training datasets. Although advanced surrogate architectures can improve predictive capability, most existing approaches fundamentally rely on increasing the quantity of high-fidelity training samples to improve model fidelity. As the physical complexity of the system increases, the amount of data required to adequately represent the underlying thermal-hydraulic behavior also increases. As a result, the additional computational cost associated with data generation may partially offset the efficiency gains provided by surrogate modeling.

Therefore, for complex full-transient thermal-hydraulic inverse uncertainty quantification problems, the primary challenge is no longer limited to accelerating Bayesian posterior exploration. An equally important challenge is reducing the number of expensive high-fidelity simulations required for surrogate construction while maintaining sufficient predictive accuracy. This issue becomes increasingly important for large-scale transient systems, where data generation may dominate the overall computational cost of the inverse uncertainty quantification process.

\subsection{Motivation for Adaptive IUQ}

To address the limitations discussed above, an adaptive inverse uncertainty quantification framework is proposed in this work. The central idea is to improve surrogate accuracy using additional high-fidelity simulations only in regions that are most relevant to Bayesian inference, rather than uniformly throughout the entire parameter space.

Conventional surrogate construction strategies typically rely on a fixed training dataset generated before Bayesian calibration begins. Since the training samples are usually distributed throughout the entire parameter domain, a significant portion of the computational effort may be spent approximating regions that have little influence on the final posterior distribution.

From the perspective of Bayesian inference, the most informative regions of the parameter space are those associated with high posterior probability. Therefore, it is desirable to allocate expensive high-fidelity simulations preferentially to these posterior-relevant regions, where improvements in surrogate accuracy can have the greatest impact on posterior estimation.

To achieve this objective, the proposed adaptive framework iteratively updates the surrogate model using information obtained from the evolving posterior distribution. New high-fidelity simulations are selectively performed in regions where surrogate uncertainty remains high and where the current posterior indicates a strong probability of future sampling.

Through this posterior-guided adaptive learning strategy, computational resources can be concentrated on the most informative regions of the parameter space, thereby improving surrogate accuracy for Bayesian inference while substantially reducing the number of expensive high-fidelity simulations required.

The resulting framework aims to maximize surrogate accuracy within posterior-relevant regions while minimizing the computational cost associated with high-fidelity data generation. The detailed adaptive inverse uncertainty quantification methodology is presented in the following section.

\section{Methodology}

The overall methodology adopted in this study is illustrated in Fig.~\ref{adaptive frame} , which combines high-fidelity simulations, PCA-GPR surrogate modeling, bayesian inverse uncertainty quantification, and adaptive posterior-guided surrogate refinement within a unified framework.

\subsection{High-fidelity thermal-hydraulic model}

The NACIE system is simulated using the Flownex SE thermal-hydraulic solver. The simulator is employed to generate the high-fidelity dataset used for surrogate construction.

The initial training dataset is defined as

\begin{equation}
D_{train}^{(0)}
=
\{
(\theta_i,y_i)
\}_{i=1}^{N_0}
\end{equation}
where $N_0$ denotes the number of initial high-fidelity simulations.

\subsection{PCA-GPR surrogate model}

To reduce the computational cost associated with repeated evaluations of
the high-fidelity simulator,
a surrogate model based on principal component analysis (PCA) and
Gaussian process regression (GPR) is constructed.
For each high-fidelity simulation, multiple transient quantities of interest are recorded throughout the entire simulation period. These quantities include the temperatures measured at different locations of the NACIE loop together with the corresponding mass flow rate. Let $N_c$ denote the number of response channels and $N_t$ denote the number of  time step.

For a given parameter realization $\theta_i$, the simulator produces a multi-output transient response

\begin{equation}
y_i
=
[y_{i,1}(t),
y_{i,2}(t),
\cdots,
y_{i,N_c}(t)]
\end{equation}
where each response channel contains a complete transient time history.

For surrogate construction, all response channels and temporal observations are concatenated into a single high-dimensional response vector,

\begin{equation}
y_i
=
[y_{i,1}(t_1),\cdots,y_{i,1}(t_{N_t}),
y_{i,2}(t_1),\cdots,y_{i,N_c}(t_{N_t})]
\end{equation}
The resulting response vector therefore contains the complete multi-output transient information generated by the high-fidelity simulation.

The simulation response matrix is written as

\begin{equation}
Y
=
[y_1,y_2,\cdots,y_{N_0}]^T
\end{equation}
PCA is applied to reduce the dimensionality of the transient response space.

The centered response matrix is

\begin{equation}
\tilde Y
=
Y-\bar Y
\end{equation}
where

\begin{equation}
\bar Y
=
\frac{1}{N_0}
\sum_{i=1}^{N_0}
y_i
\end{equation}
is the sample mean response.

The covariance matrix is

\begin{equation}
C
=
\frac{1}{N_0-1}
\tilde Y^T\tilde Y
\end{equation}
Eigenvalue decomposition gives

\begin{equation}
C
=
W\Lambda W^T
\end{equation}
where $W$ contains the principal directions and $\Lambda$ contains the
corresponding eigenvalues.

Retaining the first $K$ principal components gives the reduced PCA basis

\begin{equation}
W_K
=
[w_1,w_2,\cdots,w_K]
\end{equation}
where $W_K$ contains the first $K$ principal directions and $w_k$ denotes the $k$-th principal component direction.

For a centered transient response $\tilde y$, its projection onto the retained PCA basis is given by

\begin{equation}
z
=
W_K^T\tilde y
\end{equation}
where $W_K^T$ denotes the transpose of $W_K$. The latent variable vector is

\begin{equation}
z
=
[z_1,z_2,\cdots,z_K]
\end{equation}
where each component $z_k$ represents the coefficient of the transient response along the $k$-th principal direction. Through this projection, the original high-dimensional multi-output transient response is represented by a much lower-dimensional set of PCA coefficients.

A separate Gaussian process regression model is constructed for each
latent variable.

The predictive distribution is

\begin{equation}
z_k(\theta)
\sim
N
\left(
\mu_k(\theta),
\sigma_k^2(\theta)
\right)
\end{equation}
where $\mu_k(\theta)$ and $\sigma_k^2(\theta)$ are the predictive mean
and variance, respectively.

The original transient response is reconstructed by

\begin{equation}
\hat y
=
\bar Y
+
W_K z
\end{equation}
\subsection{Bayesian inverse uncertainty quantification}

The surrogate model is employed within a Bayesian framework to infer the posterior distribution of uncertain model parameters. In addition to the physical uncertain parameter vector $\theta$ defined previously, a model discrepancy factor $f$ is introduced to represent the relative model-form uncertainty. To ensure the positivity of $f$, the logarithm of the discrepancy factor is inferred instead of $f$ directly. Therefore, the augmented uncertain parameter vector used in Bayesian inference is written as

\begin{equation}
\phi
=
[\theta,\log f]
\end{equation}
where $\phi$ denotes the augmented parameter vector, $\theta$ denotes the physical uncertain parameter vector, and $\log f$ denotes the logarithm of the model discrepancy factor.

The prior distribution of the augmented parameter vector is defined as
\begin{equation}
p(\phi)
=
p(\theta)p(\log f),
\end{equation}
where independent prior distributions are assigned to the physical calibration parameters and the model-discrepancy parameter. The physical parameter vector is
\begin{equation}
\theta
=
[P_b,\; C_k,\; h_a,\; \lambda_p]^{\mathrm{T}},
\end{equation}

For the physical parameters, independent uniform priors are assigned over engineering-feasible ranges,
\begin{equation}
p(\theta)
=
\prod_{j=1}^{N_{\theta}} p(\theta_j),
\end{equation}
with
\begin{equation}
\theta_j
\sim
\mathcal{U}
\left(
\theta_{j,\min},
\theta_{j,\max}
\right),
\qquad
j=1,\ldots,N_{\theta}.
\end{equation}
These bounds are selected based on design information, empirical correlations, previous NACIE analyses, and expert judgment. They are intentionally broad so that the posterior distributions are not artificially constrained by overly narrow prior ranges.

The relative model-discrepancy factor $f$ is introduced to represent the residual mismatch between the model prediction and the experimental observations. Since $f$ is a positive scale parameter, the inference is performed in terms of $\log f$. A Jeffreys-type prior is assigned to $f$,
\begin{equation}
p(f)
\propto
\frac{1}{f},
\end{equation}
which is equivalent to a uniform prior in $\log f$,
\begin{equation}
\log f
\sim
\mathcal{U}
\left(
\log f_{\min},
\log f_{\max}
\right).
\end{equation}
Because the Jeffreys-type prior is improper on an unbounded domain, finite bounds are imposed on $\log f$ to obtain a proper prior distribution.

The prior bounds used in this work are summarized in Table~\ref{tab:PriorAssignment}.

\begin{table}[t!]
\centering
\caption{Prior bounds assigned to the uncertain calibration parameters used in the Bayesian IUQ analysis. The physical parameters represent heater power bias $P_b$, hydraulic resistance coefficient $C_k$, ambient heat-transfer coefficient $h_a$, and powder-gap thermal conductivity $\lambda_p$; the nuisance parameter $\log f$ controls the residual model-discrepancy scale in the likelihood.}
\begin{tabular}{l c l l l}
\hline
Parameter & Units & Min & Max & Description \\ [0.5ex]
\hline
$P_b$   & kW & $-2.5$ & $2.5$ & heater power bias \\
$C_k$   & --- & $2\cdot 10^5$ & $5\cdot 10^5$ & hydraulic resistance coefficient \\
$h_a$   & W m$^{-2}$ K$^{-1}$ & $1$ & $20$ & ambient heat-transfer coefficient \\
$\lambda_p$ & W m$^{-1}$ K$^{-1}$ & $1.5$ & $6.5$ & powder-gap thermal conductivity \\
$\log f$ & --- & $-10$ & $0$ & model-discrepancy scale factor \\ [1ex]
\hline
\end{tabular}
\label{tab:PriorAssignment}
\end{table}

The range of the heater power bias $P_b$ reflects uncertainty in heater control and power measurement. The hydraulic resistance coefficient $C_k$ accounts for lumped pressure-loss uncertainty associated with pipes, fittings, and valves. The ambient heat-transfer coefficient $h_a$ represents uncertain distributed heat loss to the surroundings, while the powder-gap thermal conductivity $\lambda_p$ captures uncertainty in the effective heat-transfer path through the heat-exchanger powder region. Posterior concentration away from the prescribed bounds indicates that the calibrated distributions are primarily informed by the NACIE measurements rather than by prior truncation.

For a given augmented parameter vector $\phi$, the surrogate prediction is obtained from the physical parameter component $\theta$ as

\begin{equation}
y_{pred}
=
M_{sur}(\theta)
\end{equation}
For a given augmented parameter vector $\phi=[\theta,\log f]$, the predicted response at observation channel $c$ and time $t_j$ is denoted by
\begin{equation}
y_{c}^{pred}(t_j;\theta)
=
M_{sur,c}(t_j;\theta),
\end{equation}
where $c=1,\ldots,N_c$ and $j=1,\ldots,N_t$. The corresponding experimental observation is denoted by $y_c^{obs}(t_j)$.

The residual between the experimental observation and the model prediction is defined as
\begin{equation}
r_{cj}(\theta)
=
y_c^{obs}(t_j)
-
y_c^{pred}(t_j;\theta).
\end{equation}

A heteroscedastic uncertainty model is adopted to account for both measurement uncertainty and residual model discrepancy. The total variance at channel $c$ and time $t_j$ is written as
\begin{equation}
\sigma_{cj}^2(\phi)
=
\sigma_{obs,cj}^2
+
\sigma_{model,cj}^2(\phi),
\end{equation}
where $\sigma_{obs,cj}$ denotes the observational uncertainty and $\sigma_{model,cj}$ denotes the model-discrepancy uncertainty.

The model-discrepancy uncertainty is represented by a relative error model,
\begin{equation}
\sigma_{model,cj}(\phi)
=
f
\left|
y_c^{pred}(t_j;\theta)
\right|,
\qquad
f=\exp(\log f),
\end{equation}
where $f>0$ controls the relative magnitude of the residual model discrepancy. This formulation allows the uncertainty level to vary with the magnitude of the transient response, which is more appropriate than a constant additive variance for multi-channel thermal-hydraulic responses with different physical scales.

Under the working assumption that the selected residuals are conditionally independent given $\phi$, the likelihood function is written as
\begin{equation}
p(D|\phi)
=
\prod_{c=1}^{N_c}
\prod_{j=1}^{N_t}
\frac{1}
{\sqrt{2\pi\sigma_{cj}^2(\phi)}}
\exp
\left[
-
\frac{
\left(
y_c^{obs}(t_j)
-
y_c^{pred}(t_j;\theta)
\right)^2
}
{
2\sigma_{cj}^2(\phi)
}
\right].
\end{equation}

Equivalently, the log-likelihood is
\begin{equation}
\log p(D|\phi)
=
-\frac{1}{2}
\sum_{c=1}^{N_c}
\sum_{j=1}^{N_t}
\left[
\log
\left(
2\pi\sigma_{cj}^2(\phi)
\right)
+
\frac{
\left(
y_c^{obs}(t_j)
-
y_c^{pred}(t_j;\theta)
\right)^2
}
{
\sigma_{cj}^2(\phi)
}
\right].
\end{equation}

The summation is taken over all selected observation channels and time steps used for Bayesian calibration. The conditional independence assumption is an approximation because transient thermal--hydraulic measurements are correlated in time and across measurement locations due to coupled conservation laws, loop inertia, and shared thermal inertia. If every available time step were included in the likelihood, these correlations would artificially inflate the effective information content of the dataset and could lead to overly narrow posterior credible intervals. To reduce this effect while retaining the dominant transient information, the original time series, containing approximately 20,000 time steps, is downsampled before calibration. Specifically, one point every 30 time steps is retained in the likelihood. This decimation reduces short-lag temporal correlation and yields a more conservative and computationally efficient calibration dataset.

According to Bayes' theorem, the posterior distribution of the augmented uncertain parameter vector is given by

\begin{equation}
p(\phi|D)
=
\frac{
p(D|\phi)p(\phi)
}
{
p(D)
}
\end{equation}
where $p(\phi|D)$ is the posterior distribution, $p(D|\phi)$ is the likelihood function, $p(\phi)$ is the prior distribution, and $p(D)$ is the evidence term. Since $p(D)$ is independent of $\phi$, the posterior distribution can be written as

\begin{equation}
p(\phi|D)
\propto
p(D|\phi)p(\phi)
\end{equation}
Posterior sampling is performed using the Microcanonical Langevin Monte Carlo (MCLMC) algorithm~\cite{robnik2024fluctuation}.

The resulting posterior samples are

\begin{equation}
\{\phi^{(s)}\}_{s=1}^{S}
\sim
p(\phi|D)
\end{equation}
where $S$ is the total number of posterior samples and $\phi^{(s)}$ denotes the $s$-th sampled augmented parameter vector from the posterior distribution. Each posterior sample contains one possible realization of the physical uncertain parameters and the model discrepancy parameter after assimilating the experimental observations.

Only the physical parameter component $\theta$ of the posterior samples is used for posterior-guided candidate generation in the adaptive refinement process, while $\log f$ is used to characterize the model discrepancy in the likelihood function.

The posterior distribution obtained in this section provides the basis for the adaptive surrogate refinement framework described in the following section.

\subsection{Adaptive posterior-guided IUQ framework}

The accuracy of surrogate-assisted inverse uncertainty quantification depends primarily on the fidelity of the surrogate in regions that make a non-negligible contribution to the posterior distribution. For a full-transient thermal-hydraulic model, enforcing uniformly high surrogate accuracy over the complete prior domain is generally inefficient because a large fraction of the prior space may have little influence on the final Bayesian inference.

An adaptive IUQ procedure is therefore developed to allocate additional high-fidelity simulations according to their expected value for posterior inference. The proposed method combines a single extensible Sobol sequence, posterior-guided local--global candidate classification, and uncertainty--diversity-based batch selection. The Sobol sequence is defined over the same fixed physical parameter domain throughout the adaptive calculation. Posterior information is used only to classify and rank the available candidates and does not translate, contract, or redefine the parameter domain.

\begin{figure}[htbp]
\centering
\includegraphics[width=\linewidth]{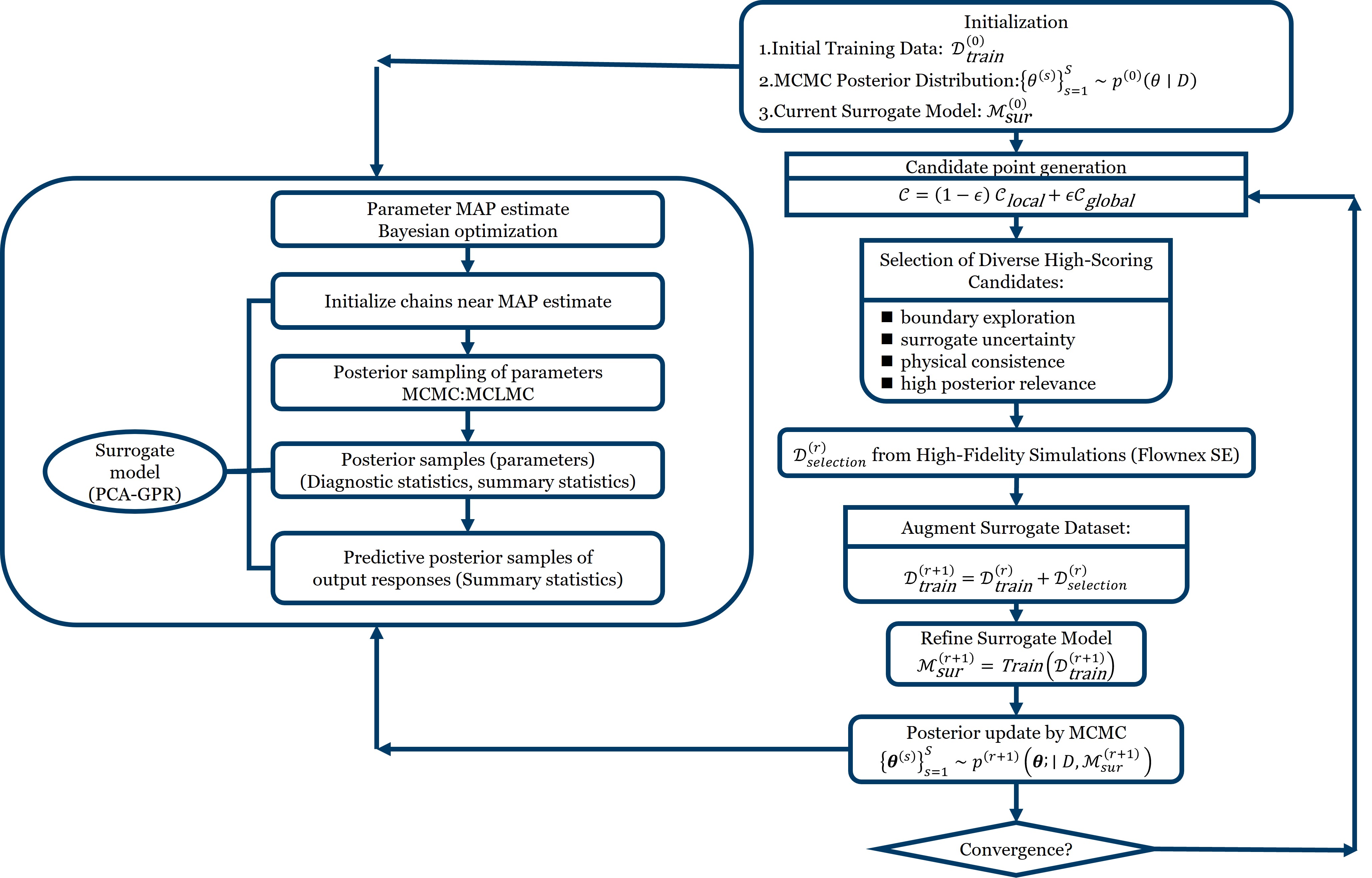}
\caption{Workflow of the extensible Sobol-sequence-based posterior-guided adaptive IUQ framework.}
\label{adaptive frame}
\end{figure}

The overall procedure is illustrated in Fig.~\ref{adaptive frame}. An initial low-discrepancy design is first generated within the prescribed prior bounds and evaluated using Flownex SE. The resulting multi-output transient responses are used to construct the initial PCA--GPR surrogate. Bayesian calibration is then performed using the surrogate, and the resulting posterior samples identify the parameter combinations that are most relevant to the experimental observations.

The first $N_0$ points of a single Sobol sequence are evaluated using Flownex SE to construct the initial high-fidelity training dataset. In the present study, $N_0=32$ is used. The appropriate initial design size is not universal and may be increased or decreased according to the dimensionality, nonlinearity, and computational complexity of the high-fidelity model. Power-of-two sample sizes, such as 32, 64, or 128, are preferable because they retain the balance properties of the Sobol construction. For models with stronger nonlinearities, more complicated transient responses, or a less smooth parameter-to-output relationship, an initial design of 64 points generally provides a more stable initial PCA basis and GPR approximation when the computational budget permits.

After the initial $N_0$ high-fidelity samples are evaluated, the same Sobol sequence is continued until a cumulative sequence length of $N_{\mathrm{Sobol},0}=1024$ is reached. The remaining unevaluated Sobol points form the initial candidate reservoir. These candidate points require only inexpensive surrogate evaluation and do not incur a Flownex SE computational cost unless they are selected for adaptive refinement.

The candidate reservoir is retained throughout the adaptive procedure. At each iteration, previously generated but unselected candidates are reclassified according to the updated posterior distribution and rescored using the updated surrogate model. If the retained reservoir contains sufficient numbers of posterior-relevant and globally exploratory candidates, no additional Sobol points are generated. If either candidate category becomes insufficient, the same Sobol sequence is extended from its current index using previously unused sequence points. The cumulative Sobol sequence length is increased in nested powers of two, for example from 1024 to 2048 and subsequently to 4096, while the original prior domain remains unchanged.

A limited batch of posterior-relevant and globally exploratory candidates is selected according to surrogate predictive uncertainty, distance from the existing training design, and within-batch diversity. Each selected parameter point is evaluated using Flownex SE and added to the high-fidelity training dataset. The unselected candidates remain in the reservoir and are reconsidered during subsequent iterations. The PCA--GPR surrogate and posterior distribution are then updated, and the procedure is repeated until the posterior statistics and representative calibrated transient responses satisfy the prescribed convergence criteria, or until the available high-fidelity simulation budget or maximum number of adaptive iterations is reached.

At adaptive iteration $r$, the available high-fidelity training dataset is defined as
\begin{equation}
\mathcal{D}_{\mathrm{train}}^{(r)}
=
\left\{
\left(
\boldsymbol{\theta}_i,
\mathbf{y}_i^{\mathrm{HF}}
\right)
\right\}_{i=1}^{N_r},
\end{equation}
where $N_r$ is the number of high-fidelity simulations available at iteration $r$, $\boldsymbol{\theta}_i$ denotes the physical calibration parameter vector, and $\mathbf{y}_i^{\mathrm{HF}}$ is the corresponding full-transient response calculated by Flownex SE.

The current PCA--GPR surrogate is written as
\begin{equation}
\mathcal{M}_{\mathrm{sur}}^{(r)}
=
\operatorname{Train}
\left(
\mathcal{D}_{\mathrm{train}}^{(r)}
\right).
\end{equation}
Bayesian IUQ is then performed using the current surrogate to obtain
\begin{equation}
p^{(r)}
\left(
\boldsymbol{\phi}
\mid
\mathcal{D}_{\mathrm{exp}},
\mathcal{M}_{\mathrm{sur}}^{(r)}
\right),
\end{equation}
where $\mathcal{D}_{\mathrm{exp}}$ denotes the experimental observations and $\boldsymbol{\phi}$ contains the complete set of inference variables. In the present study,
\begin{equation}
\boldsymbol{\phi}
=
\left[
\boldsymbol{\theta},
\log f
\right]^{\mathrm{T}},
\end{equation}
where $\boldsymbol{\theta}$ contains the physical Flownex input parameters and $\log f$ is the likelihood-error parameter. Because $\log f$ is not an input to the high-fidelity thermal-hydraulic model, only the physical parameter component $\boldsymbol{\theta}$ is used for adaptive candidate generation and selection.

Posterior samples are generated using MCLMC:
\begin{equation}
\left\{
\boldsymbol{\phi}^{(s)}
\right\}_{s=1}^{S}
\sim
p^{(r)}
\left(
\boldsymbol{\phi}
\mid
\mathcal{D}_{\mathrm{exp}},
\mathcal{M}_{\mathrm{sur}}^{(r)}
\right),
\end{equation}
where $S$ is the number of posterior samples. The corresponding physical-parameter samples
$\{\boldsymbol{\theta}^{(s)}\}_{s=1}^{S}$
are used to characterize the current posterior-relevant region and guide the subsequent adaptive refinement.

\subsubsection{Extensible Sobol candidate reservoir}

Let the physically admissible parameter domain be
\begin{equation}
\Omega
=
\prod_{j=1}^{d}
\left[
\theta_{j,\min},
\theta_{j,\max}
\right],
\end{equation}
where $d$ is the number of physical calibration parameters.

A single extensible Sobol sequence is defined in the unit hypercube:
\begin{equation}
\mathcal{S}
=
\left\{
\mathbf{u}_1,
\mathbf{u}_2,
\ldots
\right\},
\qquad
\mathbf{u}_i
\in
[0,1]^d.
\end{equation}

Each normalized Sobol point is mapped to the physical parameter domain using the fixed affine transformation
\begin{equation}
\boldsymbol{\theta}_i^{\mathrm{Sobol}}
=
\boldsymbol{\theta}_{\min}
+
\mathbf{u}_i
\odot
\left(
\boldsymbol{\theta}_{\max}
-
\boldsymbol{\theta}_{\min}
\right),
\end{equation}
where $\odot$ denotes component-wise multiplication.

The same lower and upper bounds are retained throughout the adaptive calculation. Consequently, all initial and subsequently generated candidates belong to the same parameter domain $\Omega$. The adaptive process does not shift the normalized interval away from $[0,1]^d$, nor does it shrink the physical parameter bounds according to the current posterior distribution.

The initial high-fidelity training design is constructed from the first $N_0$ points of the Sobol sequence:
\begin{equation}
\mathcal{X}_{\mathrm{init}}
=
\left\{
\boldsymbol{\theta}_i^{\mathrm{Sobol}}
\right\}_{i=1}^{N_0}.
\end{equation}

In the present application, $N_0=32$. However, $N_0$ may be adjusted according to the complexity of the high-fidelity response. A relatively simple and smooth model may be adequately represented by a smaller initial design, whereas a highly nonlinear or strongly coupled model may require 64 or more initial high-fidelity samples. Whenever practical, a power-of-two value is recommended so that the initial design corresponds to a balanced Sobol prefix.

The same sequence is then continued to a prescribed initial cumulative length $n_0$:
\begin{equation}
n_0
=
N_{\mathrm{Sobol},0},
\end{equation}
where $N_{\mathrm{Sobol},0}=1024$ in the present study. After excluding the first $N_0$ high-fidelity training points, the initial unevaluated candidate reservoir is
\begin{equation}
\mathcal{R}_{\mathrm{ret}}^{(0)}
=
\left\{
\boldsymbol{\theta}_i^{\mathrm{Sobol}}
\right\}_{i=N_0+1}^{n_0}.
\end{equation}
Thus, for $N_0=32$ and $n_0=1024$, the initial reservoir contains 992 unevaluated candidates. If $N_0=64$ is used instead, the same 1024-point Sobol prefix produces an initial reservoir of 960 unevaluated candidates.

At iteration $r$, let $\mathcal{R}_{\mathrm{ret}}^{(r)}$ denote the candidates retained from the preceding iterations. These candidates have not yet been selected for high-fidelity evaluation. They are first reclassified using the current posterior distribution.

If the retained reservoir satisfies the prescribed candidate-number requirements, the Sobol sequence is not extended:
\begin{equation}
n_r
=
n_{r-1}.
\end{equation}
If either the local or global candidate subset is insufficient, the cumulative Sobol sequence length is increased to the next nested level:
\begin{equation}
n_r
=
\min
\left(
2n_{r-1},
n_{\max}
\right),
\end{equation}
where $n_{\max}$ is the maximum allowable Sobol prefix length, in the present application, $n_{\max}=32768$.

The newly appended candidate block is
\begin{equation}
\mathcal{C}_{\mathrm{new}}^{(r)}
=
\left\{
\boldsymbol{\theta}_i^{\mathrm{Sobol}}
\right\}_{i=n_{r-1}+1}^{n_r}.
\end{equation}

When no Sobol extension is required,
\begin{equation}
\mathcal{C}_{\mathrm{new}}^{(r)}
=
\varnothing.
\end{equation}

The complete candidate reservoir available for classification and selection at iteration $r$ is therefore
\begin{equation}
\mathcal{R}^{(r)}
=
\mathcal{R}_{\mathrm{ret}}^{(r)}
\cup
\mathcal{C}_{\mathrm{new}}^{(r)}.
\end{equation}
The sequence is always continued from its current index. Consequently, extending the candidate design from 1024 to 2048 points adds only the previously unused Sobol points with indices 1025--2048. It does not restart the sequence or regenerate an independent Sobol design.

\subsubsection{Posterior-guided local--global partition}

All physical parameters are expressed in the normalized coordinate system associated with the fixed Sobol domain:
\begin{equation}
\widetilde{\theta}_j
=
\frac{
\theta_j-\theta_{j,\min}
}{
\theta_{j,\max}-\theta_{j,\min}
},
\qquad
j=1,\ldots,d.
\end{equation}

The empirical mean and covariance matrix of the normalized posterior samples at iteration $r$ are denoted by
\begin{equation}
\widetilde{\boldsymbol{\mu}}_{\mathrm{post}}^{(r)}
=
\frac{1}{S}
\sum_{s=1}^{S}
\widetilde{\boldsymbol{\theta}}^{(s)}
\end{equation}
and
\begin{equation}
\widetilde{\boldsymbol{\Sigma}}_{\mathrm{post}}^{(r)}
=
\operatorname{Cov}
\left(
\left\{
\widetilde{\boldsymbol{\theta}}^{(s)}
\right\}_{s=1}^{S}
\right).
\end{equation}

When the posterior is strongly concentrated or contains highly correlated parameters, the empirical covariance matrix may become poorly conditioned. A regularized covariance matrix is therefore used:
\begin{equation}
\widetilde{\boldsymbol{\Sigma}}_{\mathrm{reg}}^{(r)}
=
\widetilde{\boldsymbol{\Sigma}}_{\mathrm{post}}^{(r)}
+
\lambda_{\Sigma}
\frac{
\operatorname{tr}
\left(
\widetilde{\boldsymbol{\Sigma}}_{\mathrm{post}}^{(r)}
\right)
}{d}
\mathbf{I},
\end{equation}
where $\lambda_{\Sigma}$ is a small positive regularization coefficient and $\mathbf{I}$ is the identity matrix.

For any candidate $\boldsymbol{\theta}$ contained in the current reservoir, its squared Mahalanobis distance from the posterior sample cloud is defined as
\begin{equation}
D_M^2
\left(
\widetilde{\boldsymbol{\theta}}
\right)
=
\left(
\widetilde{\boldsymbol{\theta}}
-
\widetilde{\boldsymbol{\mu}}_{\mathrm{post}}^{(r)}
\right)^{\mathrm{T}}
\left(
\widetilde{\boldsymbol{\Sigma}}_{\mathrm{reg}}^{(r)}
\right)^{-1}
\left(
\widetilde{\boldsymbol{\theta}}
-
\widetilde{\boldsymbol{\mu}}_{\mathrm{post}}^{(r)}
\right).
\end{equation}

The same distance is evaluated for every normalized posterior sample:
\begin{equation}
D_{M,s}^2
=
D_M^2
\left(
\widetilde{\boldsymbol{\theta}}^{(s)}
\right),
\qquad
s=1,\ldots,S.
\end{equation}

An empirical posterior-support threshold is defined as
\begin{equation}
c_p^{(r)}
=
Q_p
\left(
\left\{
D_{M,s}^2
\right\}_{s=1}^{S}
\right),
\end{equation}
where $Q_p$ denotes the empirical $p$-quantile of the posterior-sample Mahalanobis distances. In the present study, $p=0.95$ is used.

All candidates in the current reservoir are classified into the posterior-relevant local subset
\begin{equation}
\mathcal{R}_{\mathrm{local}}^{(r)}
=
\left\{
\boldsymbol{\theta}
\in
\mathcal{R}^{(r)}
:
D_M^2
\left(
\widetilde{\boldsymbol{\theta}}
\right)
\leq
c_p^{(r)}
\right\}
\end{equation}
and the globally exploratory subset
\begin{equation}
\mathcal{R}_{\mathrm{global}}^{(r)}
=
\mathcal{R}^{(r)}
\setminus
\mathcal{R}_{\mathrm{local}}^{(r)}.
\end{equation}

Because the posterior distribution changes after each surrogate refinement, candidates retained from previous iterations may change classification. A previously global candidate may become posterior-relevant at a later iteration, while a previously local candidate may move outside the updated posterior-support region. The complete reservoir is therefore reclassified at every adaptive iteration.

This classification does not alter the Sobol sampling domain. It determines how the limited high-fidelity simulation budget is allocated within the fixed global candidate design.

The number of new high-fidelity evaluations at iteration $r$ is
\begin{equation}
B
=
B_{\mathrm{local}}
+
B_{\mathrm{global}}.
\end{equation}

A prescribed fraction of the batch is retained for global exploration:
\begin{equation}
B_{\mathrm{global}}
=
\max
\left[
1,
\operatorname{round}
\left(
\epsilon_B B
\right)
\right],
\end{equation}
and
\begin{equation}
B_{\mathrm{local}}
=
B-B_{\mathrm{global}},
\end{equation}
where $\epsilon_B$ is the global exploration fraction.

In the present study, $\epsilon_B=0.25$ is used. A batch containing eight new high-fidelity simulations therefore contains six posterior-relevant points and two globally exploratory points. If either candidate subset contains fewer points than its prescribed batch allocation, the unfilled portion is transferred to the other subset.

To ensure that the acquisition procedure retains sufficient competition among candidates, the minimum required local and global reservoir sizes are defined as
\begin{equation}
N_{\mathrm{local,min}}
=
\gamma_{\mathrm{local}}
B_{\mathrm{local}}
\end{equation}
and
\begin{equation}
N_{\mathrm{global,min}}
=
\gamma_{\mathrm{global}}
B_{\mathrm{global}},
\end{equation}
where $\gamma_{\mathrm{local}}$ and $\gamma_{\mathrm{global}}$ are candidate oversampling factors.

In the present implementation,
\begin{equation}
\gamma_{\mathrm{local}}
=
\gamma_{\mathrm{global}}
=
10.
\end{equation}

For $B=8$ and $\epsilon_B=0.25$, the corresponding minimum candidate requirements are
\begin{equation}
N_{\mathrm{local,min}}
=
60,
\qquad
N_{\mathrm{global,min}}
=
20.
\end{equation}

The Sobol sequence is extended only if
\begin{equation}
\left|
\mathcal{R}_{\mathrm{local}}^{(r)}
\right|
<
N_{\mathrm{local,min}}
\end{equation}
or
\begin{equation}
\left|
\mathcal{R}_{\mathrm{global}}^{(r)}
\right|
<
N_{\mathrm{global,min}}.
\end{equation}

If both requirements are satisfied, the retained candidate reservoir is used directly and no new Sobol candidates are generated during that iteration.

\subsubsection{Uncertainty--diversity candidate selection}

At each adaptive iteration, all candidates in the current reservoir are evaluated using the updated PCA--GPR surrogate. Previously generated but unselected candidates are therefore rescored after every surrogate and posterior update.

For a candidate parameter vector $\boldsymbol{\theta}$, the predictive distribution of the retained PCA coefficients is written as
\begin{equation}
\mathbf{z}
\left(
\boldsymbol{\theta}
\right)
\sim
\mathcal{N}
\left(
\boldsymbol{\mu}_z
\left(
\boldsymbol{\theta}
\right),
\boldsymbol{\Sigma}_z
\left(
\boldsymbol{\theta}
\right)
\right).
\end{equation}

A scalar surrogate uncertainty indicator is defined as
\begin{equation}
u
\left(
\boldsymbol{\theta}
\right)
=
\frac{1}{K}
\sum_{k=1}^{K}
\frac{
\sigma_k^2
\left(
\boldsymbol{\theta}
\right)
}{
s_{z_k}^2+\varepsilon
},
\end{equation}
where $\sigma_k^2(\boldsymbol{\theta})$ is the GPR predictive variance of the $k$-th PCA coefficient, $s_{z_k}^2$ is the variance of the corresponding coefficient in the current training dataset, $K$ is the number of retained PCA components, and $\varepsilon$ is a small positive constant. When the PCA coefficients are standardized before GPR training, $s_{z_k}^2=1$.

Candidate novelty is quantified by the minimum distance from the current training design:
\begin{equation}
d_{\mathrm{train}}^{(r)}
\left(
\boldsymbol{\theta}
\right)
=
\min_{
\boldsymbol{\theta}_i
\in
\mathcal{X}_{\mathrm{train}}^{(r)}
}
\left\|
\widetilde{\boldsymbol{\theta}}
-
\widetilde{\boldsymbol{\theta}}_i
\right\|_2.
\end{equation}

All distances are evaluated in the normalized parameter space. This prevents parameters with large physical magnitudes or broad numerical ranges from dominating the geometric selection criteria.

The uncertainty and training-distance indicators are normalized separately within the local and global candidate subsets:
\begin{equation}
\widetilde{u}
\left(
\boldsymbol{\theta}
\right)
=
\frac{
u(\boldsymbol{\theta})-u_{\min}
}{
u_{\max}-u_{\min}+\varepsilon
}
\end{equation}
and
\begin{equation}
\widetilde{d}_{\mathrm{train}}^{(r)}
\left(
\boldsymbol{\theta}
\right)
=
\frac{
d_{\mathrm{train}}^{(r)}(\boldsymbol{\theta})-d_{\min}
}{
d_{\max}-d_{\min}+\varepsilon
}.
\end{equation}

Separate normalization is used because local and global candidates are selected under separate batch allocations. If the range of either metric is smaller than a prescribed numerical tolerance, the corresponding normalized quantity is set to unity for every candidate in that subset so that a non-informative metric does not affect their relative ranking.

The base acquisition score is defined as
\begin{equation}
A_{\mathrm{base}}^{(r)}
\left(
\boldsymbol{\theta}
\right)
=
\left[
\widetilde{u}
\left(
\boldsymbol{\theta}
\right)
\right]^{\alpha_u}
\left[
\widetilde{d}_{\mathrm{train}}^{(r)}
\left(
\boldsymbol{\theta}
\right)
\right]^{\beta_d},
\end{equation}
where $\alpha_u$ and $\beta_d$ control the relative emphasis placed on surrogate uncertainty and geometric novelty. In the present implementation,
\begin{equation}
\alpha_u
=
\beta_d
=
1.
\end{equation}

The adaptive batch is constructed sequentially to prevent multiple candidates from clustering in the same region. Before selecting the $m$-th point, the minimum distance between a candidate and the points already selected in the current batch is calculated as
\begin{equation}
d_{\mathrm{batch},m}^{(r)}
\left(
\boldsymbol{\theta}
\right)
=
\min_{
\boldsymbol{\theta}'
\in
\mathcal{B}_{m-1}^{(r)}
}
\left\|
\widetilde{\boldsymbol{\theta}}
-
\widetilde{\boldsymbol{\theta}}'
\right\|_2,
\end{equation}
where $\mathcal{B}_{m-1}^{(r)}$ contains all local and global points selected before the $m$-th selection at iteration $r$. For the first selected point,
\begin{equation}
d_{\mathrm{batch},1}^{(r)}
=
1.
\end{equation}

The sequential acquisition score is
\begin{equation}
A_m^{(r)}
\left(
\boldsymbol{\theta}
\right)
=
A_{\mathrm{base}}^{(r)}
\left(
\boldsymbol{\theta}
\right)
\left[
d_{\mathrm{batch},m}^{(r)}
\left(
\boldsymbol{\theta}
\right)
\right]^{\lambda_d},
\end{equation}
where $\lambda_d=1$ in the present implementation.

Local and global candidates are selected according to their prescribed allocations. At each selection step, the candidate with the largest sequential acquisition score is added to the batch, after which the distances to the selected batch are updated. The resulting design favors candidates associated with high surrogate uncertainty and insufficient existing coverage while avoiding redundant high-fidelity simulations within the same adaptive iteration.

\subsubsection{High-fidelity evaluation and surrogate refinement}

The selected adaptive batch is written as
\begin{equation}
\mathcal{B}^{(r)}
=
\mathcal{B}_{\mathrm{local}}^{(r)}
\cup
\mathcal{B}_{\mathrm{global}}^{(r)}.
\end{equation}

Every selected parameter point is evaluated using a new Flownex SE simulation:
\begin{equation}
\mathcal{D}_{\mathrm{selection}}^{(r)}
=
\left\{
\left(
\boldsymbol{\theta}_k,
\mathbf{y}_k^{\mathrm{HF}}
\right)
\right\}_{k=1}^{B}.
\end{equation}

Only the selected candidates incur a high-fidelity computational cost. All other candidates remain unevaluated and are retained for possible selection during subsequent adaptive iterations.

The surrogate training dataset is augmented according to
\begin{equation}
\mathcal{D}_{\mathrm{train}}^{(r+1)}
=
\mathcal{D}_{\mathrm{train}}^{(r)}
\cup
\mathcal{D}_{\mathrm{selection}}^{(r)}.
\end{equation}

The selected candidates are removed from the reservoir, whereas all unselected candidates are retained:
\begin{equation}
\mathcal{R}_{\mathrm{ret}}^{(r+1)}
=
\mathcal{R}^{(r)}
\setminus
\mathcal{B}^{(r)}.
\end{equation}

The PCA basis and the associated GPR models are retrained using the complete augmented dataset, while the retained latent dimension $K$ is kept fixed. The refined surrogate is then used to update the posterior distribution:
\begin{equation}
p^{(r+1)}
\left(
\boldsymbol{\phi}
\mid
\mathcal{D}_{\mathrm{exp}},
\mathcal{M}_{\mathrm{sur}}^{(r+1)}
\right).
\end{equation}

The updated posterior samples are subsequently used to reclassify and rescore the retained candidate reservoir. The Sobol sequence is extended only when the retained reservoir no longer satisfies the prescribed local or global candidate-number requirements.

\subsubsection{Convergence assessment}

Convergence is assessed from the stability of the posterior parameter statistics and the representative calibrated transient response. Let
$\boldsymbol{\mu}_{\theta}^{(r)}$
and
$\boldsymbol{\sigma}_{\theta}^{(r)}$
denote the posterior mean and posterior standard deviation of the physical calibration parameters at iteration $r$.

The normalized change in posterior mean is defined as
\begin{equation}
\Delta_{\mu}^{(r)}
=
\frac{1}{\sqrt{d}}
\left\|
\frac{
\boldsymbol{\mu}_{\theta}^{(r)}
-
\boldsymbol{\mu}_{\theta}^{(r-1)}
}{
\boldsymbol{\theta}_{\max}
-
\boldsymbol{\theta}_{\min}
}
\right\|_2,
\end{equation}
and the normalized change in posterior standard deviation is
\begin{equation}
\Delta_{\sigma}^{(r)}
=
\frac{1}{\sqrt{d}}
\left\|
\frac{
\boldsymbol{\sigma}_{\theta}^{(r)}
-
\boldsymbol{\sigma}_{\theta}^{(r-1)}
}{
\boldsymbol{\theta}_{\max}
-
\boldsymbol{\theta}_{\min}
}
\right\|_2.
\end{equation}

The complete transient response evaluated at the current posterior mean is
\begin{equation}
\mathbf{y}_{\mathrm{mean}}^{(r)}
=
\mathcal{M}_{\mathrm{sur}}^{(r)}
\left(
\boldsymbol{\mu}_{\theta}^{(r)}
\right).
\end{equation}

Its relative change between two consecutive iterations is
\begin{equation}
\Delta_y^{(r)}
=
\frac{
\left\|
\mathbf{y}_{\mathrm{mean}}^{(r)}
-
\mathbf{y}_{\mathrm{mean}}^{(r-1)}
\right\|_2
}{
\left\|
\mathbf{y}_{\mathrm{mean}}^{(r-1)}
\right\|_2
+
\varepsilon
}.
\end{equation}

The average surrogate uncertainty over the current posterior samples is also monitored:
\begin{equation}
\overline{u}^{(r)}
=
\frac{1}{S}
\sum_{s=1}^{S}
u
\left(
\boldsymbol{\theta}^{(s)}
\right),
\end{equation}
with the corresponding relative change
\begin{equation}
\Delta_u^{(r)}
=
\frac{
\left|
\overline{u}^{(r)}
-
\overline{u}^{(r-1)}
\right|
}{
\overline{u}^{(r-1)}
+
\varepsilon
}.
\end{equation}

Because the PCA basis and GPR models are retrained as the dataset is augmented, $\Delta_u^{(r)}$ is used as an auxiliary diagnostic rather than as the sole termination criterion.

The adaptive procedure is considered converged when
\begin{equation}
\Delta_{\mu}^{(r)}
<
\tau_{\mu},
\qquad
\Delta_{\sigma}^{(r)}
<
\tau_{\sigma},
\qquad
\Delta_y^{(r)}
<
\tau_y
\end{equation}
for two consecutive adaptive iterations.

In the present study,
\begin{equation}
\tau_{\mu}
=
0.05,
\qquad
\tau_{\sigma}
=
0.05,
\qquad
\tau_y
=
0.01,
\end{equation}
while
\begin{equation}
\tau_u
=
0.01
\end{equation}
is used for the auxiliary surrogate-uncertainty diagnostic.

The adaptive refinement is terminated when the posterior-statistical and calibrated-response convergence conditions are satisfied for two consecutive iterations. The calculation is also terminated if the prescribed high-fidelity simulation budget or the maximum number of adaptive iterations is reached before convergence.

Unlike Bayesian optimization, the objective of the present framework is not to locate a single optimal parameter vector. The convergence criteria therefore measure the stability of the inferred posterior distribution and the corresponding calibrated transient responses rather than convergence toward a point optimum.

\section{Results }

This section presents a comprehensive analysis of the convergence behavior, predictive performance, and computational efficiency of the proposed adaptive IUQ framework. The results demonstrate the capability of the framework to provide reliable posterior inference and efficient surrogate refinement for complex transient thermal-hydraulic responses.

\subsection{Convergence behavior and batch-size robustness of adaptive IUQ}

Based on the prior ranges listed in Table~\ref{tab:PriorAssignment}\cite{YAO2026140531}, a relatively small initial training set containing 32 samples was generated using Sobol sampling. Starting from this initial dataset, adaptive IUQ was then performed to progressively refine the surrogate model and update the posterior inference. To assess the robustness and stability of the proposed adaptive IUQ framework, two different batch sizes were considered, in which either 8 or 16 new candidate points were added at each adaptive iteration. This comparison also shows the influence of the number of newly added high-fidelity samples per iteration on the convergence behavior to be examined.

\begin{figure}[h]
\centering
\includegraphics[width=1\linewidth]{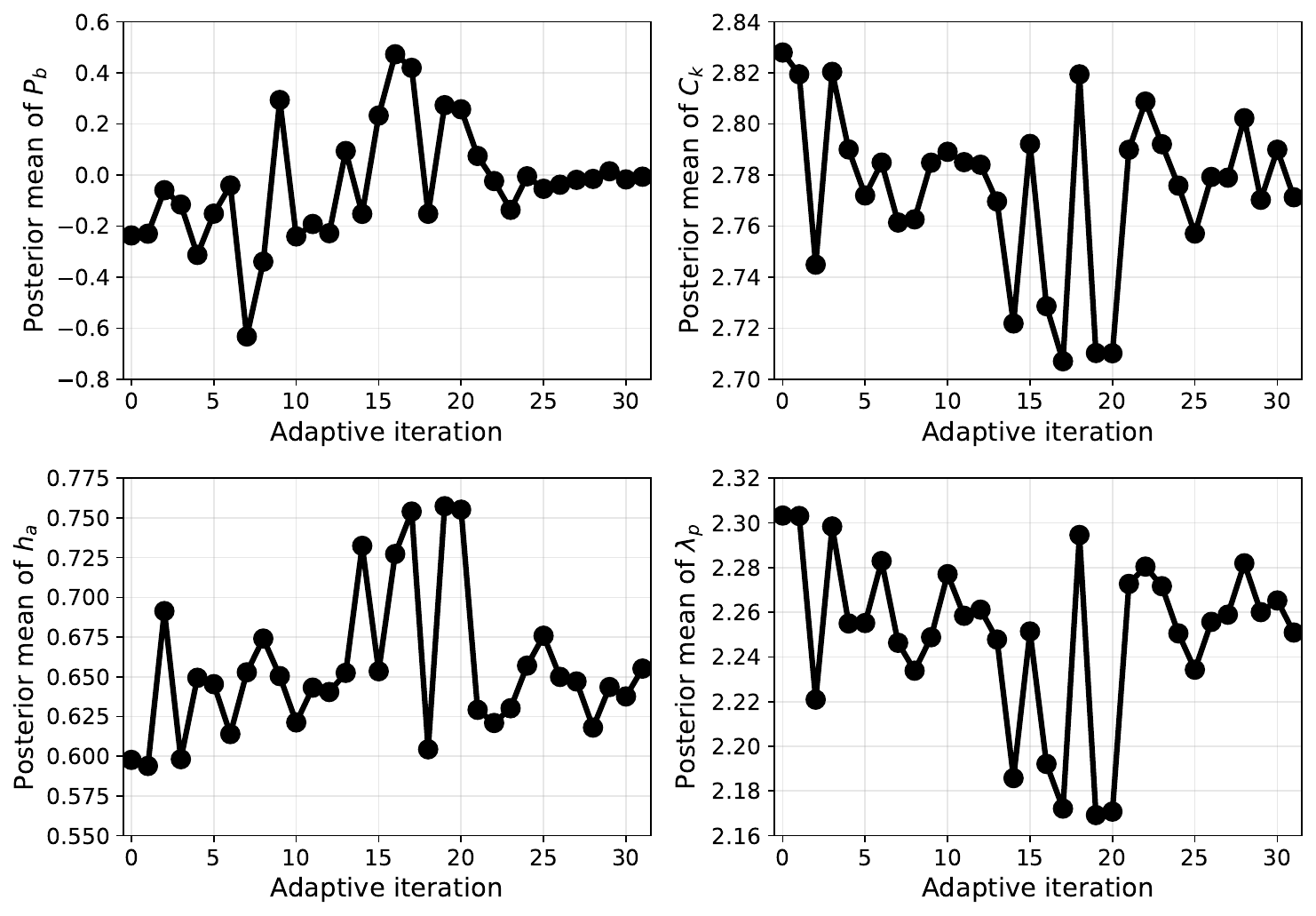}
\caption{Evolution of posterior means of uncertain parameters during adaptive iterations at 8 selected candidate points }
\label{fig:convergence-8}
\end{figure}

\begin{figure}[h]
\centering
\includegraphics[width=1\linewidth]{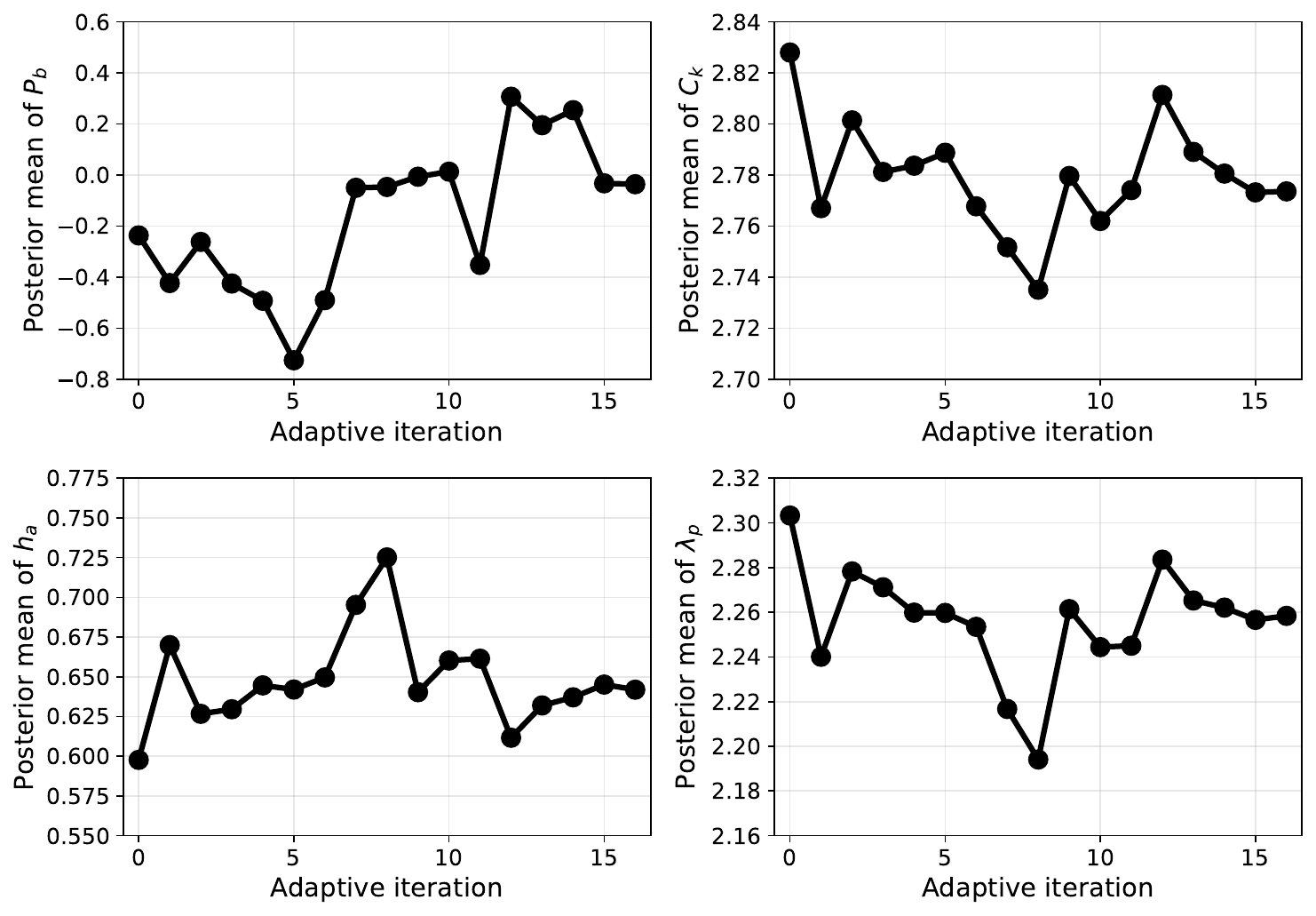}
\caption{Evolution of posterior means of uncertain parameters during adaptive iterations at 16 selected candidate points}
\label{fig:convergence-16}
\end{figure}

Figure~\ref{fig:convergence-8} shows the evolution of the posterior means when 8 new candidate points are added at each adaptive iteration, while Fig.~\ref{fig:convergence-16} presents the corresponding results when 16 new candidate points are added per iteration. Although different batch sizes lead to slightly different convergence trajectories, the overall convergence trends are highly consistent. More importantly, the final converged posterior mean values are nearly identical for the two settings. This indicates that the adaptive IUQ results are not sensitive to the specific batch-size setting, demonstrating the robustness and general applicability of the proposed framework.

It is also worth noting that, when 8 new samples are added at each iteration, the convergence criterion is satisfied after 31 adaptive iterations, corresponding to a total of 248 newly added high-fidelity samples. In comparison, when 16 new samples are added at each iteration, convergence is achieved after 16 adaptive iterations, corresponding to a total of 256 newly added high-fidelity samples. Therefore, despite the different numbers of samples added per iteration, the total number of additional high-fidelity simulations required for convergence remains at a similar level. This further confirms that the proposed adaptive IUQ framework exhibits stable convergence behavior and robust performance under different refinement settings.    

\subsection{Posterior inference accuracy and validation of the adaptive IUQ framework}

The posterior sampling in the adaptive IUQ framework was performed using the MCLMC algorithm. For each adaptive iteration, the PCA-GPR surrogate model constructed from the current training dataset was embedded into the Bayesian IUQ formulation, and MCLMC was used to generate posterior samples of the uncertain parameters. The convergence and mixing behavior of the Markov chains were assessed using trace plots and autocorrelation function (ACF) plots, following the convergence diagnostics adopted in Yao et al.~\cite{yao1}. In particular, well-mixed trace plots without long-term drift and rapidly decaying ACF values were taken as indicators of satisfactory posterior sampling performance.
\begin{figure}[h]
    \centering

    \begin{subfigure}[b]{0.45\linewidth}
        \centering
        \includegraphics[width=\linewidth]{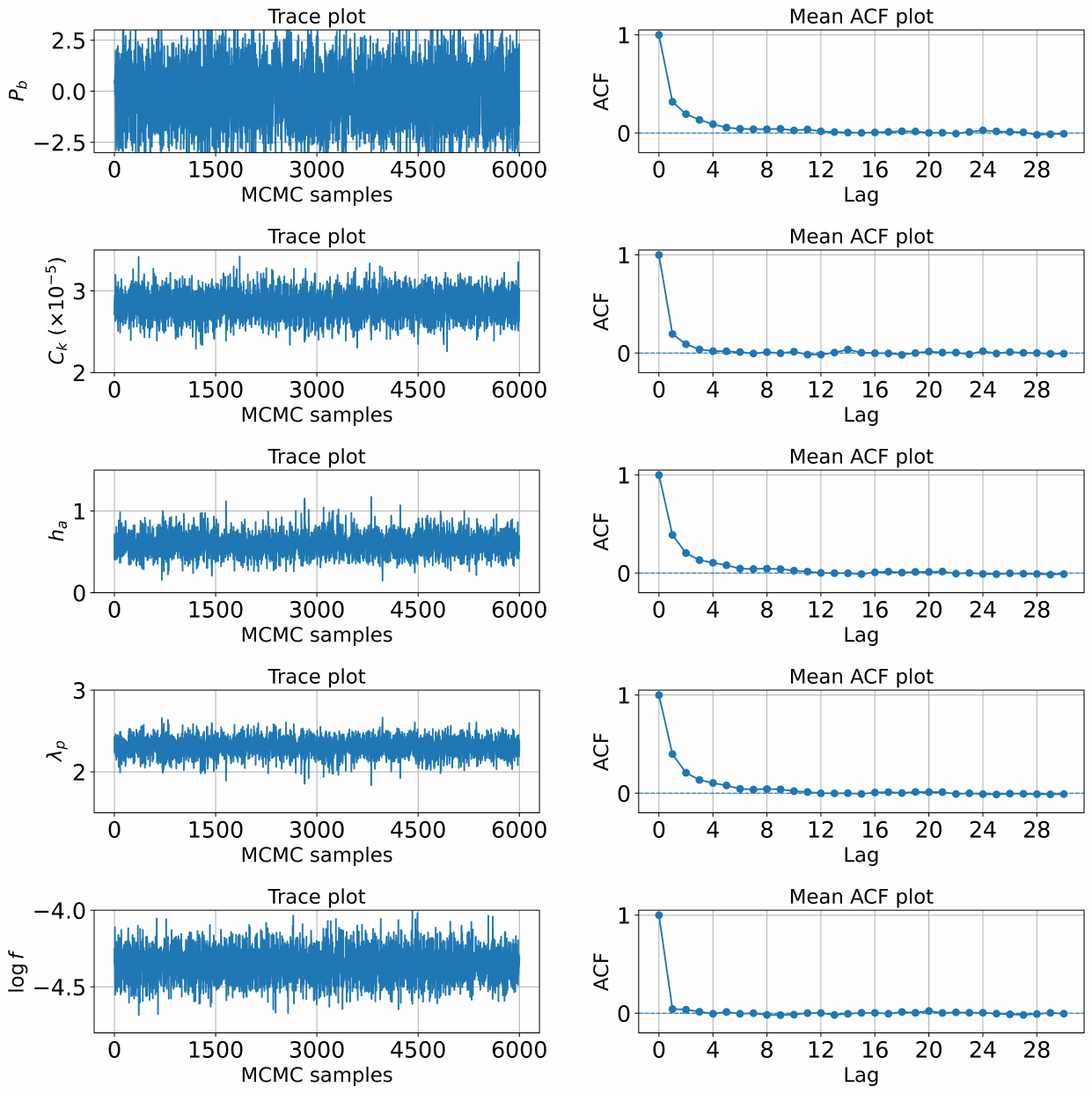}
        \caption{Trace plots and ACF plots obtained for various components of \(\bm{\theta}\) at first iteration.}
        \label{fig:acf32-0}
    \end{subfigure}
    \hfill
    \begin{subfigure}[b]{0.45\linewidth}
        \centering
        \includegraphics[width=\linewidth]{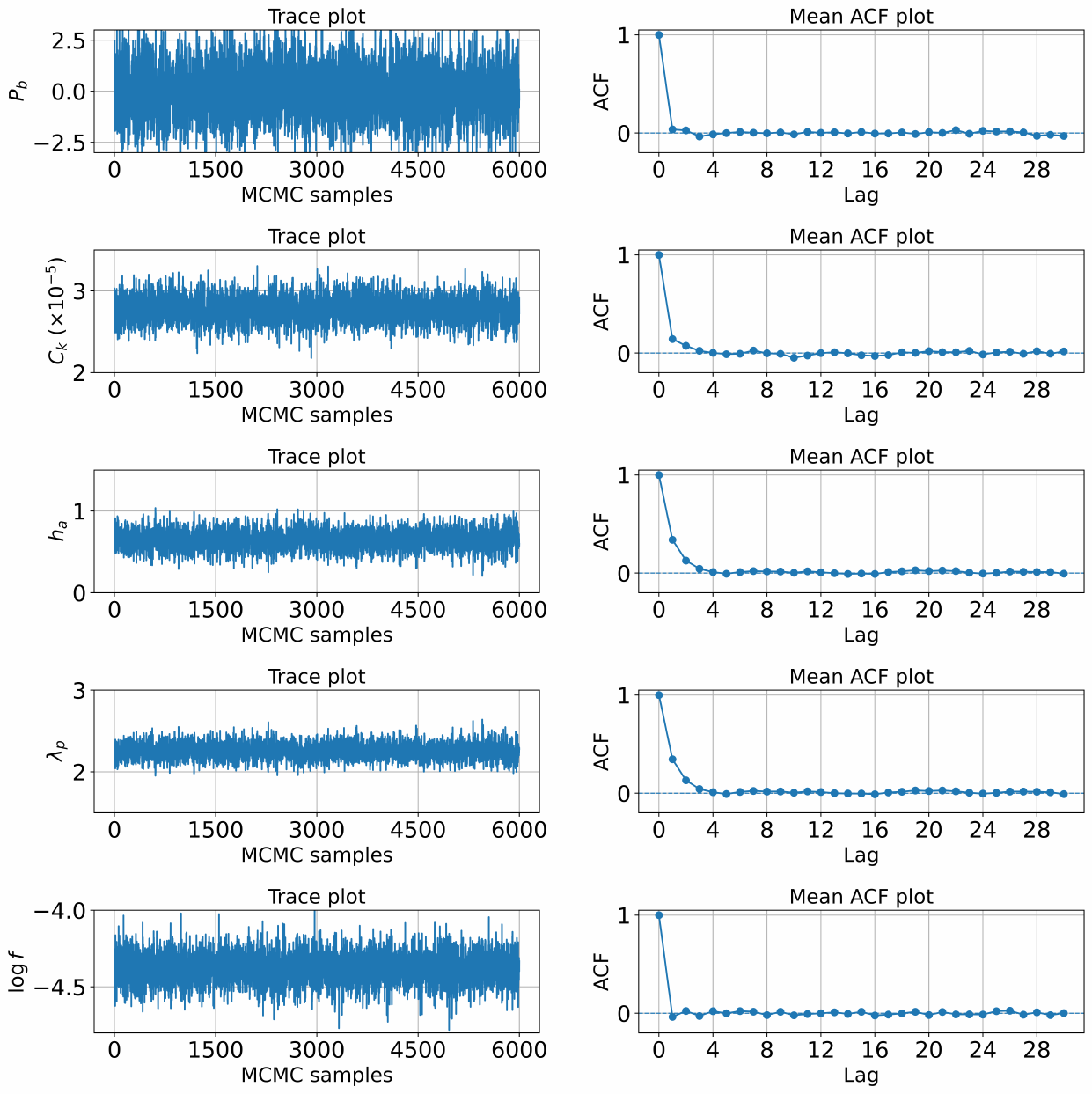}
        \caption{Trace plots and ACF plots obtained for various components of \(\bm{\theta}\) at 15th iteration.}
        \label{fig:acf32-16}
    \end{subfigure}
    \caption{Trace plots for various components of \(\bm{\theta}\) and autocorrelation function plots for the corresponding parameter samples produced at different iterations. In each subfigure, the left-hand panels show the trace plots and the right-hand panels show the ACF plots.}
    \label{fig:iterations_comparison_acf}
\end{figure}

\begin{figure}[h]
    \centering

    \begin{subfigure}[b]{0.45\linewidth}
        \centering
        \includegraphics[width=\linewidth]{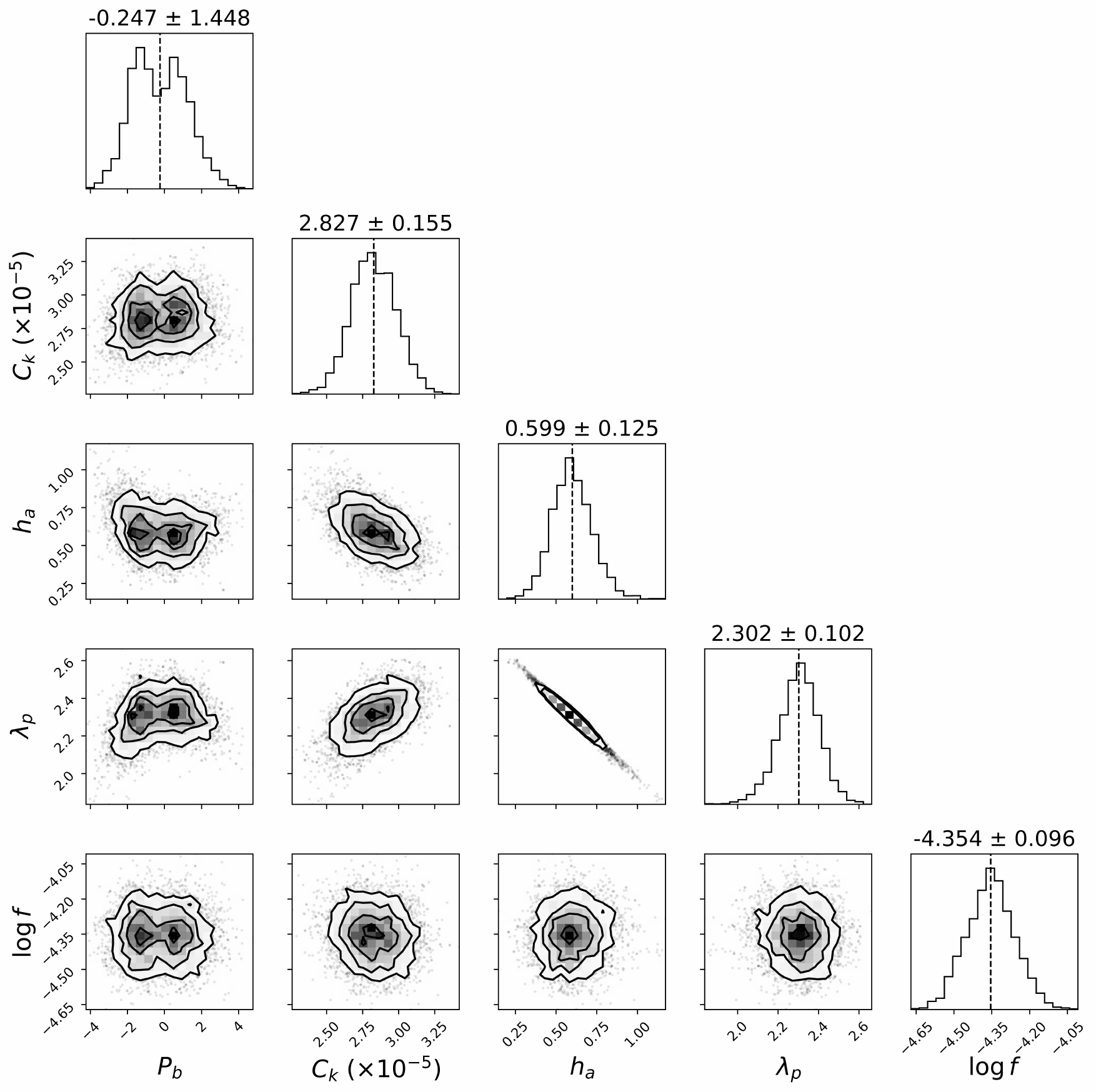}
        \caption{Marginal posterior distributions of the parameters sampled at first iteration.}
        \label{fig:post32-0}
    \end{subfigure}
    \hfill
    \begin{subfigure}[b]{0.45\linewidth}
        \centering
        \includegraphics[width=\linewidth]{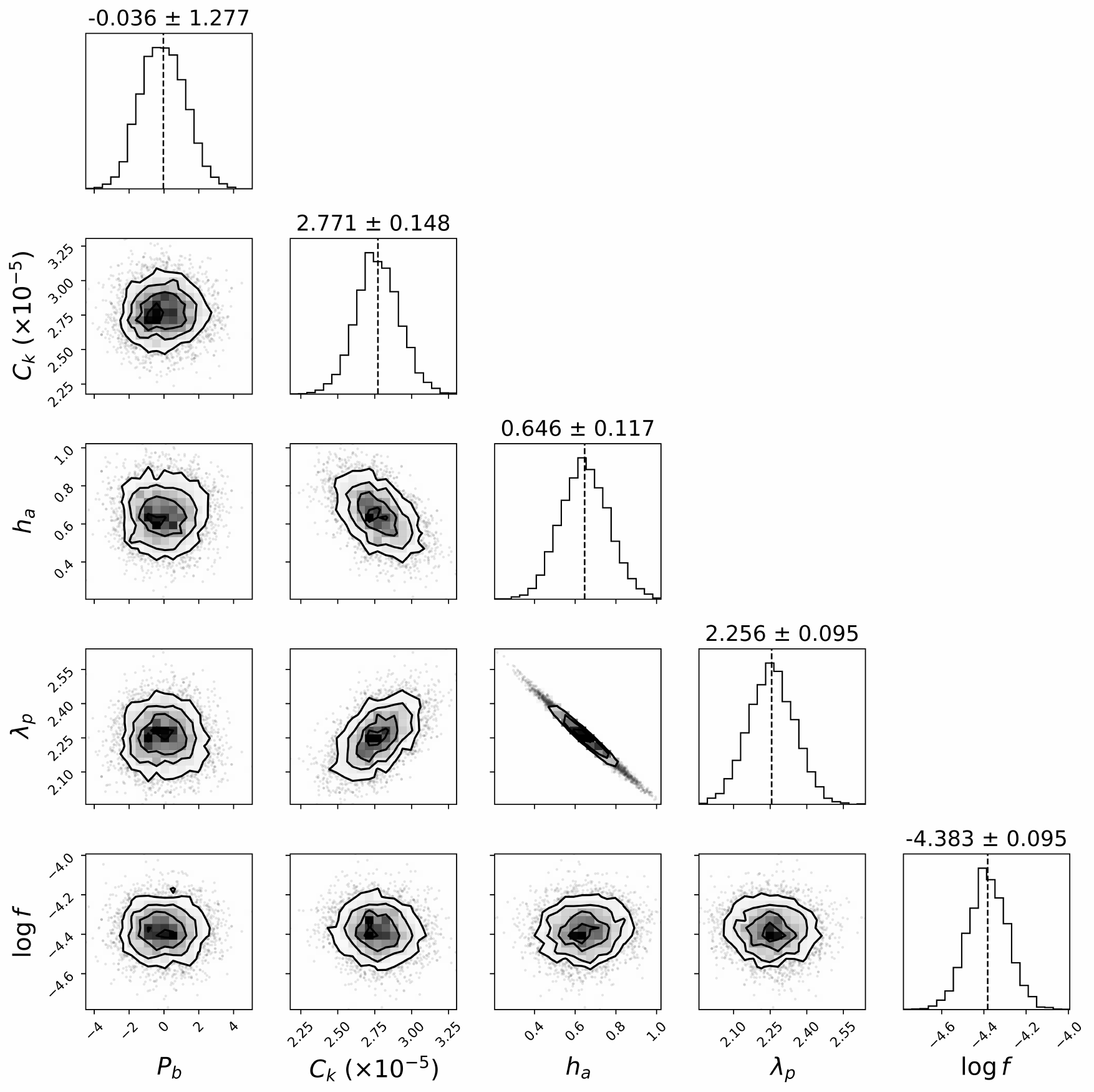}
        \caption{Marginal posterior distributions of the parameters sampled at 15th iteration.}
        \label{fig:post32-16}
    \end{subfigure}
    \caption{Univariate diagonal and bivariate off-diagonal marginal posterior distributions of the parameters, namely the components of the parameter vector \(\bm{\theta}\), obtained from samples drawn from the posterior distribution \(p(\bm{\theta}\mid D,I)\)  at different iterations.}
    \label{fig:iterations_comparison_pos}
\end{figure}

Figure~\ref{fig:iterations_comparison_acf} compares the trace plots and ACF plots obtained at different adaptive iterations. For both the initial iteration and the later adaptive iteration, the trace plots of the uncertain parameters fluctuate around statistically stationary regions without persistent trends or systematic drift. This indicates that the MCLMC sampler is able to explore the posterior distribution effectively under both surrogate configurations. In addition, the corresponding ACF curves decay rapidly with increasing lag, suggesting weak sample correlation and efficient exploration of the posterior space. These results confirm that the posterior samples obtained at each selected adaptive iteration are statistically reliable and can be used to characterize the inferred parameter distributions.

Although the MCLMC chains exhibit satisfactory convergence at both iterations, the resulting posterior distributions are not identical, as shown in Fig.~\ref{fig:iterations_comparison_pos}. This difference is expected because the posterior distribution inferred in surrogate-assisted IUQ depends not only on the experimental observations and prior information, but also on the accuracy of the surrogate model in the posterior-relevant region. At the initial iteration, the PCA-GPR surrogate is constructed from a relatively limited training dataset generated by Sobol sampling. Although this initial surrogate provides a feasible approximation for Bayesian inference, its local accuracy in the high-probability posterior region may still be insufficient. As a result, the inferred posterior distribution may exhibit biased mean values, overly broad or narrow marginal distributions, or less accurate correlation structures between parameters.

As the adaptive IUQ procedure progresses, additional high-fidelity simulations are selected and added to the training dataset according to posterior relevance and surrogate uncertainty. This targeted enrichment improves the accuracy of the PCA-GPR surrogate in the regions that have the greatest influence on Bayesian calibration. Consequently, the posterior distributions obtained at later iterations become more representative of the high-fidelity model response. As shown in Fig.~\ref{fig:iterations_comparison_pos}, the marginal posterior distributions and the bivariate correlation structures evolve from the initial iteration to the later iteration. The shifts in posterior means and changes in posterior variances indicate that the adaptive refinement process corrects the inference results as the surrogate model becomes more accurate in the posterior region.

It should be emphasized that the posterior distributions at both iterations are converged from the MCMC sampling perspective. Therefore, the differences observed in the posterior distributions should not be interpreted as sampling failure. Instead, they reflect the influence of surrogate-model accuracy on Bayesian inference. The adaptive IUQ framework progressively reduces this surrogate-induced inference error by allocating new high-fidelity simulations to informative regions of the parameter space. This behavior demonstrates a key advantage of the proposed method: rather than relying on a fixed surrogate constructed from a limited initial design, the posterior distribution is iteratively updated and refined as the surrogate accuracy improves.

Overall, the trace and ACF diagnostics verify the reliability of the MCLMC sampling process, while the comparison of posterior distributions demonstrates the role of adaptive surrogate refinement in improving posterior inference. The evolution of the posterior distribution toward a more stable and physically consistent region confirms that the proposed adaptive IUQ framework can effectively couple efficient posterior sampling with targeted surrogate improvement. This provides strong evidence that the framework is suitable for complex transient thermal-hydraulic systems, where direct high-fidelity Bayesian calibration would be computationally prohibitive.

\begin{figure}[h]
    \centering

    \begin{subfigure}[b]{0.45\linewidth}
        \centering
        \includegraphics[width=\linewidth]{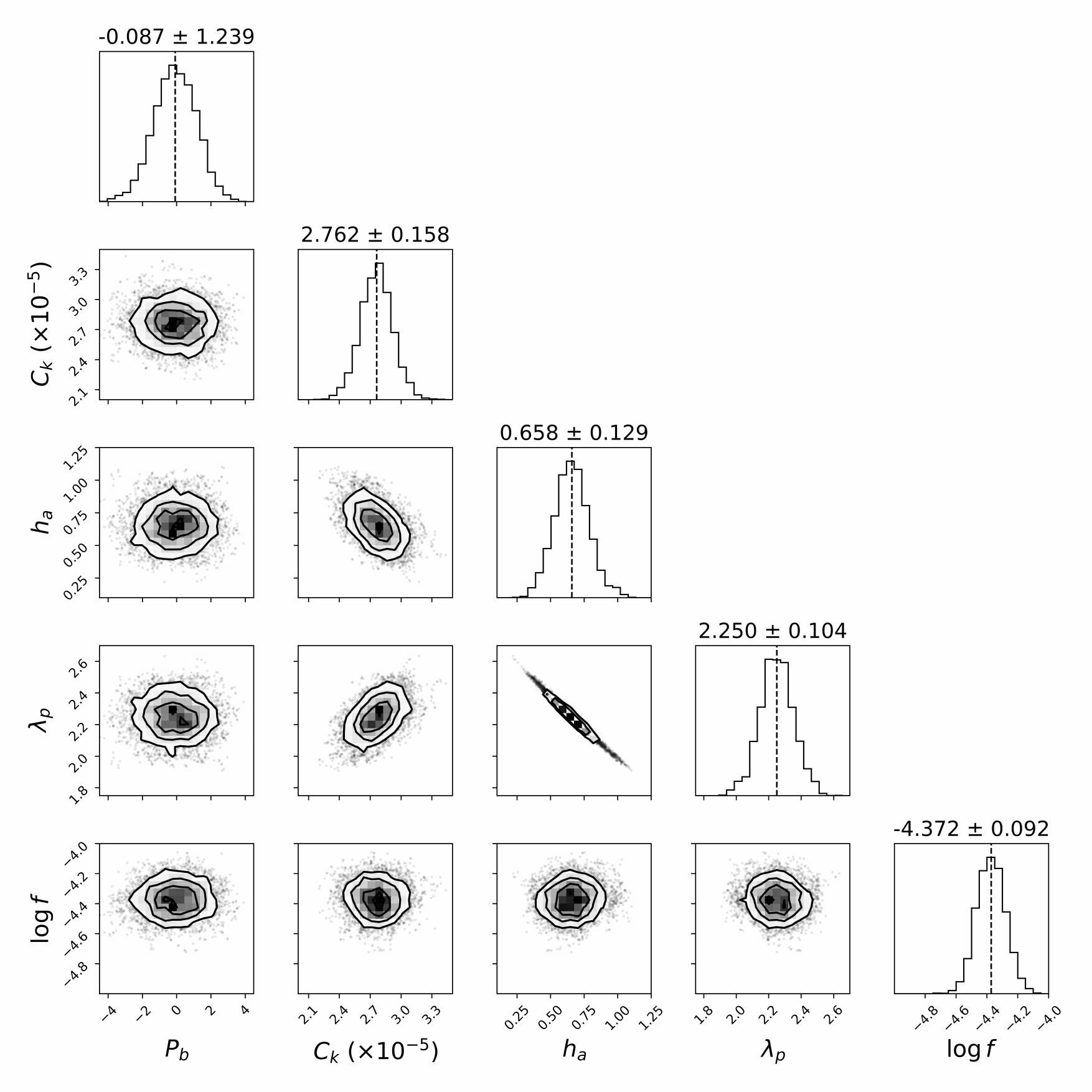}
        \caption{Marginal posterior distributions of the parameters sampled with Flownex SE.}
        \label{fig:postflownex}
    \end{subfigure}
    \hfill
    \begin{subfigure}[b]{0.45\linewidth}
        \centering
        \includegraphics[width=\linewidth]{pic/post32-15.pdf}
        \caption{Marginal posterior distributions of the parameters sampled with adaptive IUQ.}
        \label{fig:post32-16-com}
    \end{subfigure}
    \caption{Comparison of univariate diagonal and bivariate off-diagonal marginal posterior distributions obtained from the direct Flownex-based IUQ and the proposed adaptive IUQ framework. The direct Flownex-based posterior is used as the high-fidelity reference, while the adaptive IUQ posterior is obtained using the refined PCA-GPR surrogate.}
    \label{fig:iterations_comparison_flownex}
\end{figure}
Figure~\ref{fig:iterations_comparison_flownex} compares the posterior distributions obtained from the direct Flownex-based IUQ and the proposed adaptive IUQ framework. The direct Flownex-based posterior is used as the high-fidelity reference because the likelihood evaluations are performed using the original Flownex SE model. In contrast, the adaptive IUQ posterior is obtained using the refined PCA-GPR surrogate after adaptive enrichment. Therefore, this comparison provides a direct assessment of whether the adaptive surrogate can reproduce the posterior inference obtained from the high-fidelity model. As shown in Fig.~\ref{fig:iterations_comparison_flownex}, the posterior distributions obtained by the two approaches exhibit a high degree of consistency. The marginal posterior distributions show similar peak locations and comparable widths for all uncertain parameters, indicating that both the central posterior estimates and the posterior uncertainty levels are accurately captured by the adaptive IUQ framework. In addition, the bivariate posterior distributions demonstrate that the refined PCA-GPR surrogate preserves the joint posterior structure of the high-fidelity Flownex-based inference. This is particularly evident from the strong nonlinear dependence between the ambient heat transfer coefficient $h_a$ and the powder gap thermal conductivity $\lambda_p$, which is consistently reproduced by the adaptive IUQ result. Therefore, the proposed framework can capture not only the marginal posterior behavior of individual parameters, but also the nonlinear parameter correlations that arise from coupled thermal-hydraulic effects.

The visual agreement observed in Fig.~\ref{fig:iterations_comparison_flownex} is further supported by the quantitative posterior statistics reported in Table~\ref{tab:posterior_summary_three_methods}. The posterior means, standard deviations, medians, and selected percentiles obtained from adaptive IUQ are close to those obtained from the direct Flownex-based IUQ. For example, the posterior means of $C_k$, $h_a$, $\lambda_p$, and $\log(f)$ are nearly identical between the two approaches, and their posterior standard deviations are also of the same magnitude. The percentile ranges, including $P_5$ to $P_{95}$, further confirm that the adaptive IUQ framework preserves the main uncertainty characteristics of the Flownex-based posterior.

For completeness, the relative errors of the posterior mean and posterior standard deviation are summarized in Table~\ref{tab:posterior_error_flownex_mclmc}. For the $j$-th parameter, the relative error of the posterior mean is defined as

\begin{equation}
E_{\mu,j}
=
\frac{
|\mu_{j}^{MCLMC}-\mu_{j}^{Flownex}|
}
{
|\mu_{j}^{Flownex}|+\epsilon
}
\times 100\%
\end{equation}

and the relative error of the posterior standard deviation is defined as

\begin{equation}
E_{\sigma,j}
=
\frac{
|\sigma_{j}^{MCLMC}-\sigma_{j}^{Flownex}|
}
{
|\sigma_{j}^{Flownex}|+\epsilon
}
\times 100\%
\end{equation}

where $\mu_{j}^{Flownex}$ and $\sigma_{j}^{Flownex}$ denote the posterior mean and standard deviation obtained from the direct Flownex-based IUQ, while $\mu_{j}^{MCLMC}$ and $\sigma_{j}^{MCLMC}$ denote the corresponding quantities obtained from the adaptive IUQ with MCLMC. The small constant $\epsilon$ is introduced to avoid numerical instability when the reference value is close to zero.
\begin{table}[htbp]
\centering
\caption{Posterior summary statistics obtained from the direct Flownex-based IUQ, fixed-surrogate MCLMC, and adaptive IUQ with MCLMC.}
\label{tab:posterior_summary_three_methods}

\resizebox{\textwidth}{!}{%
\begin{tabular}{
    l
    S[table-format=-1.6]
    S[table-format=1.6]
    S[table-format=-1.6]
    S[table-format=-1.6]
    S[table-format=-1.6]
    S[table-format=1.6]
    S[table-format=1.6]
}
\toprule
\multicolumn{8}{c}{Power bias $P_b$} \\
\midrule
{Method} & {Mean} & {Std} & {P5} & {P25} & {P50} & {P75} & {P95} \\
\midrule
Flownex model         & -0.087153 & 1.238632 & -2.142147 & -0.911214 & -0.077645 & 0.768213 & 1.900745 \\
Fixed surrogate MCLMC & -0.059069 & 1.278880 & -2.165022 & -0.943309 & -0.066537 & 0.828459 & 2.043015 \\
Adaptive IUQ MCLMC    & -0.032734 & 1.292366 & -2.092823 & -0.920613 & -0.055331 & 0.828720 & 2.104373 \\
\midrule

\multicolumn{8}{c}{Resistance coefficient $C_k$($\times 10^{-5}$)} \\
\midrule
{Method} & {Mean} & {Std} & {P5} & {P25} & {P50} & {P75} & {P95} \\
\midrule
Flownex model         & 2.762034 & 0.157742 & 2.503568 & 2.657221 & 2.762652 & 2.863036 & 3.024897 \\
Fixed surrogate MCLMC & 2.779015 & 0.152141 & 2.532866 & 2.674258 & 2.778158 & 2.880997 & 3.032645 \\
Adaptive IUQ MCLMC    & 2.773178 & 0.149899 & 2.530200 & 2.674646 & 2.770757 & 2.871299 & 3.021383 \\
\midrule

\multicolumn{8}{c}{Ambient heat transfer coefficient $h_a$($\times 10$)} \\
\midrule
{Method} & {Mean} & {Std} & {P5} & {P25} & {P50} & {P75} & {P95} \\
\midrule
Flownex model         & 0.657812 & 0.129198 & 0.447495 & 0.572203 & 0.654229 & 0.738791 & 0.871575 \\
Fixed surrogate MCLMC & 0.648384 & 0.120352 & 0.449348 & 0.567101 & 0.648745 & 0.730117 & 0.842230 \\
Adaptive IUQ MCLMC    & 0.645106 & 0.118112 & 0.455135 & 0.566712 & 0.643850 & 0.722837 & 0.843060 \\
\midrule

\multicolumn{8}{c}{Powder gap thermal conductivity $\lambda_p$} \\
\midrule
{Method} & {Mean} & {Std} & {P5} & {P25} & {P50} & {P75} & {P95} \\
\midrule
Flownex model         & 2.249572 & 0.104243 & 2.081149 & 2.181709 & 2.249384 & 2.317133 & 2.424116 \\
Fixed surrogate MCLMC & 2.253750 & 0.097590 & 2.101297 & 2.185883 & 2.251510 & 2.318146 & 2.419717 \\
Adaptive IUQ MCLMC    & 2.256601 & 0.096366 & 2.098892 & 2.192173 & 2.256214 & 2.319970 & 2.414391 \\
\midrule

\multicolumn{8}{c}{Logarithm of model discrepancy factor $\log(f)$} \\
\midrule
{Method} & {Mean} & {Std} & {P5} & {P25} & {P50} & {P75} & {P95} \\
\midrule
Flownex model         & -4.372476 & 0.092369 & -4.518862 & -4.437001 & -4.373734 & -4.309006 & -4.220075 \\
Fixed surrogate MCLMC & -4.379469 & 0.095007 & -4.534247 & -4.444808 & -4.379811 & -4.314889 & -4.223755 \\
Adaptive IUQ MCLMC    & -4.382811 & 0.095312 & -4.538502 & -4.445676 & -4.382613 & -4.318875 & -4.227051 \\
\bottomrule
\end{tabular}
}
\end{table}

As shown in Table~\ref{tab:posterior_error_flownex_mclmc}, the relative errors for most posterior means and standard deviations remain small, confirming the quantitative agreement between the two approaches. The relatively large mean error observed for $P_b$ is mainly caused by the small magnitude of the reference posterior mean. In absolute terms, the difference between the two posterior means of $P_b$ remains small compared with the corresponding posterior standard deviation, and therefore does not indicate a significant discrepancy in posterior inference.

Overall, the comparison between Fig.~\ref{fig:iterations_comparison_flownex}, Table~\ref{tab:posterior_summary_three_methods}, and Table~\ref{tab:posterior_error_flownex_mclmc} demonstrates that the proposed adaptive IUQ framework can accurately reproduce the posterior distribution obtained from direct Flownex-based IUQ. This confirms that the adaptive refinement strategy successfully improves the PCA-GPR surrogate in the posterior-relevant region. As a result, the final adaptive surrogate can retain the accuracy of high-fidelity Bayesian inference while substantially reducing the number of direct Flownex SE evaluations required for posterior sampling.

\begin{table}[htbp]
\centering
\caption{Relative differences of posterior mean and standard deviation between adaptive IUQ with MCLMC and direct Flownex-based IUQ.}
\label{tab:posterior_error_flownex_mclmc}

\resizebox{\textwidth}{!}{%
\begin{tabular}{
    l
    S[table-format=-1.6]
    S[table-format=-1.6]
    S[table-format=2.3]
    S[table-format=1.6]
    S[table-format=1.6]
    S[table-format=2.3]
}
\toprule
{Parameter} 
& {Flownex Mean} 
& {MCLMC Mean} 
& {Mean Error (\%)} 
& {Flownex Std} 
& {MCLMC Std} 
& {Std Error (\%)} \\
\midrule
Power bias $P_b$ 
& -0.087153 
& -0.032734 
& 62.441 
& 1.238632 
& 1.292366 
& 4.338 \\

Resistance coefficient $C_k$ 
& 2.762034 
& 2.773178 
& 0.403 
& 0.157742 
& 0.149899 
& 4.972 \\

Ambient heat transfer coefficient $h_a$ 
& 0.657812 
& 0.645106 
& 1.932 
& 0.129198 
& 0.118112 
& 8.581 \\

Powder gap thermal conductivity $\lambda_p$ 
& 2.249572 
& 2.256601 
& 0.312 
& 0.104243 
& 0.096366 
& 7.556 \\

Logarithm of model discrepancy factor $\log(f)$ 
& -4.372476 
& -4.382811 
& 0.236 
& 0.092369 
& 0.095312 
& 3.186 \\
\bottomrule
\end{tabular}
}
\end{table}

\subsection{Posterior Parameter--Response Associations and System-Level Physical Interpretation}

The preceding posterior and predictive comparisons demonstrate that the adaptively refined PCA--GPR surrogate can reproduce the marginal parameter distributions, joint posterior dependence structures, and calibrated transient responses obtained from direct Flownex-based IUQ. To further interpret the engineering significance of the calibrated posterior, the uncertain physical parameters are related to a set of derived system-level indicators describing the natural-circulation and heat-removal behavior of the NACIE loop. The four calibrated physical parameters represent distinct but strongly coupled thermal--hydraulic mechanisms. The heater power bias $P_b$ characterizes uncertainty in the imposed thermal input and primarily affects the overall energy supplied to the loop. The hydraulic resistance coefficient $C_k$ represents the lumped pressure losses associated with the loop components and therefore directly influences the balance between the buoyancy driving head and the total hydraulic resistance. The ambient heat-transfer coefficient $h_a$ controls the distributed heat loss from the loop to the surrounding environment. The powder-gap thermal conductivity $\lambda_p$ governs the effective thermal resistance of the heat-exchanger powder region and consequently affects the transfer of thermal energy from the primary coolant to the secondary cooling side.

For each posterior parameter sample $\boldsymbol{\theta}^{(s)}$, the complete multi-channel transient response was reconstructed using the final adaptive PCA--GPR surrogate. Four derived indicators were then extracted to characterize the system-level natural-circulation and heat-removal behavior. The time-averaged indicators were evaluated over the final 10\% of the transient response, corresponding to the interval $18000\leq t\leq19980~\mathrm{s}$ for the present dataset. This interval is referred to as the terminal analysis window and is used to characterize the late-stage system response.

The terminal-window mean mass flow rate was defined as
\begin{equation}
\overline{\dot{m}}_{\mathrm{term}}
=
\frac{1}{N_{\mathrm{term}}}
\sum_{n\in\mathcal{I}_{\mathrm{term}}}
\left|\dot{m}_n\right|,
\end{equation}
where
\begin{equation}
\mathcal{I}_{\mathrm{term}}
=
\left\{
n\,\middle|\,18000\leq t_n\leq19980~\mathrm{s}
\right\},
\qquad
N_{\mathrm{term}}=198.
\end{equation}

The maximum heater-section outlet temperature was evaluated over the complete transient as
\begin{equation}
T_{\mathrm{HS,out}}^{\max}
=
\max_{1\leq n\leq N_t}
T_{\mathrm{HS,out},n},
\end{equation}
where $N_t=2001$ is the total number of retained time points.

The terminal-window mean heat-exchanger temperature difference was calculated as
\begin{equation}
\overline{\Delta T}_{\mathrm{HX,term}}
=
\frac{1}{N_{\mathrm{term}}}
\sum_{n\in\mathcal{I}_{\mathrm{term}}}
\left(
T_{\mathrm{HX,in},n}
-
T_{\mathrm{HX,out},n}
\right).
\end{equation}

The corresponding terminal-window mean heat-removal rate was evaluated as
\begin{equation}
\overline{\dot{Q}}_{\mathrm{HX,term}}
=
\frac{1}{N_{\mathrm{term}}}
\sum_{n\in\mathcal{I}_{\mathrm{term}}}
\left|\dot{m}_n\right|c_{p,n}
\left(
T_{\mathrm{HX,in},n}
-
T_{\mathrm{HX,out},n}
\right).
\end{equation}

The association between calibration parameter $\theta_j$ and system-level indicator $q_k$ was quantified using the Spearman rank correlation coefficient,
\begin{equation}
\rho_{jk}^{\mathrm{S}}
=
\operatorname{Corr}
\left[
\operatorname{rank}(\theta_j),
\operatorname{rank}(q_k)
\right].
\end{equation}
Spearman correlation was adopted because the parameter--response relationships within the posterior-supported region may be monotonic but nonlinear. The resulting coefficients describe associations among the experimentally constrained posterior samples and should not be interpreted as global variance-based sensitivity indices or as isolated causal effects of individual parameters. In particular, because the calibrated parameters exhibit posterior dependence, a large coefficient may reflect both the direct physical effect of a parameter and its coupled variation with other parameters.

\begin{figure}[htbp]
\centering
\includegraphics[width=1\textwidth]{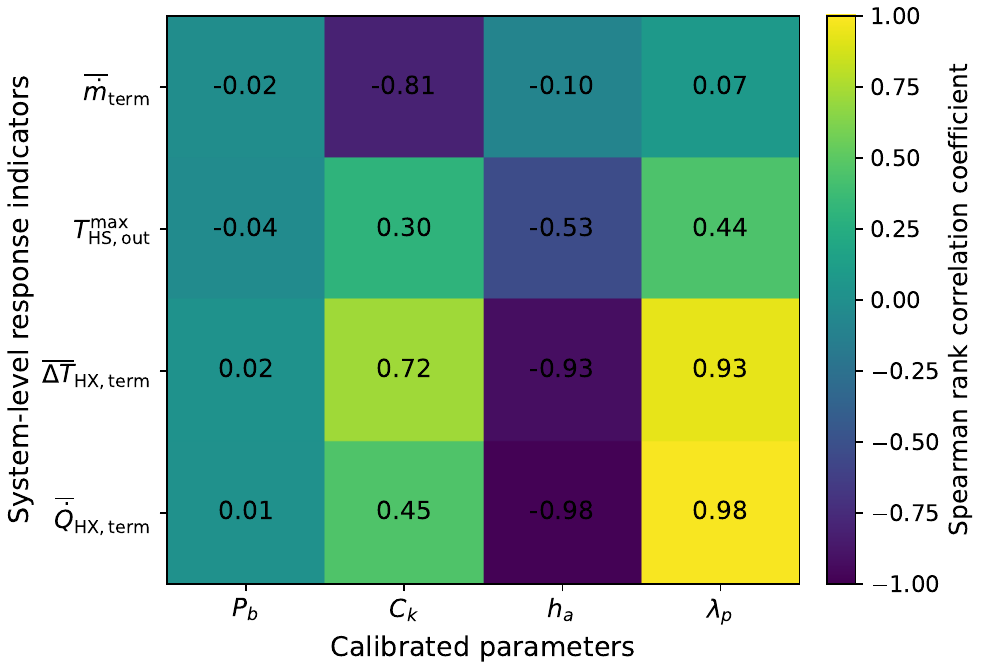}
\caption{Spearman rank correlations between the calibrated physical parameters and the derived system-level response indicators evaluated over the posterior distribution.}
\label{fig:ss}
\end{figure}

As shown in Fig.~\ref{fig:ss}, the posterior association structure reveals two distinct controlling mechanisms in the NACIE system. The terminal-window mean natural-circulation mass flow rate is primarily governed by the hydraulic resistance coefficient $C_k$. The strong negative correlation,
\begin{equation}
\rho_{\mathrm{S}}
\left(
C_k,\overline{\dot{m}}_{\mathrm{term}}
\right)
=
-0.81,
\end{equation}
is consistent with the fundamental pressure-balance mechanism of a natural-circulation loop. An increase in $C_k$ raises the distributed hydraulic pressure loss and consequently reduces the flow rate that can be sustained by the available buoyancy driving head. By comparison, the posterior associations of $\overline{\dot{m}}_{\mathrm{term}}$ with $P_b$, $h_a$, and $\lambda_p$ are weak. This result indicates that, after conditioning on the experimental observations, the remaining variation in the terminal-window mean mass flow rate is dominated by hydraulic resistance rather than by direct variation in the thermal parameters.

The maximum heater-section outlet temperature exhibits a more strongly coupled dependence structure. Its moderate negative association with $h_a$, $\rho_{\mathrm{S}}=-0.53$, is physically consistent with the reduction in the loop temperature level caused by stronger distributed heat loss to the surroundings. The moderate positive association with $C_k$, $\rho_{\mathrm{S}}=0.30$, is consistent with the reduction in convective energy transport and the increase in fluid residence time resulting from a lower circulation rate. The positive association between $\lambda_p$ and $T_{\mathrm{HS,out}}^{\max}$, $\rho_{\mathrm{S}}=0.44$, reflects the system-wide propagation of the heat-exchanger thermal resistance. Although $\lambda_p$ is defined for the powder-gap region of the heat exchanger, its influence is not confined to the local heat-exchanger temperature difference. An increase in $\lambda_p$ reduces the effective powder-gap thermal resistance and alters the heat-removal rate, the heat-exchanger outlet temperature, and consequently the temperature of the coolant recirculating toward the heater section. These changes propagate through the entire natural-circulation loop by modifying the bulk temperature distribution, the buoyancy driving head, the circulation rate, and the transient accumulation of thermal energy. The maximum heater-section outlet temperature is therefore a system-level response that is naturally affected by the heat-exchanger performance.

The positive sign of the posterior association represents the net response along the experimentally constrained parameter manifold rather than the isolated effect of changing $\lambda_p$ while holding all other parameters fixed. As shown in Fig.~\ref{fig:post32-16}, $\lambda_p$ is strongly coupled with $h_a$. Posterior samples with larger $\lambda_p$ are accompanied by compensating changes in distributed ambient heat loss, which alter the partition of energy between the heat exchanger and the surrounding environment. Consequently, the enhanced heat-exchanger conductance associated with a larger $\lambda_p$ does not act independently; its local cooling effect is combined with the simultaneous change in the overall heat-loss distribution. The observed positive correlation therefore represents the resulting system-level balance among heat-exchanger removal, ambient heat dissipation, natural-circulation transport, and thermal-energy storage.

The heat-exchanger indicators exhibit the strongest posterior associations. The terminal-window mean heat-exchanger temperature difference satisfies
\begin{equation}
\rho_{\mathrm{S}}
\left(
h_a,\overline{\Delta T}_{\mathrm{HX,term}}
\right)
=
-0.93,
\qquad
\rho_{\mathrm{S}}
\left(
\lambda_p,\overline{\Delta T}_{\mathrm{HX,term}}
\right)
=
0.93,
\end{equation}
while the terminal-window mean heat-removal rate satisfies
\begin{equation}
\rho_{\mathrm{S}}
\left(
h_a,\overline{\dot{Q}}_{\mathrm{HX,term}}
\right)
=
-0.98,
\qquad
\rho_{\mathrm{S}}
\left(
\lambda_p,\overline{\dot{Q}}_{\mathrm{HX,term}}
\right)
=
0.98.
\end{equation}
These nearly opposite associations reflect competition between two system-level heat-removal pathways. Increasing $h_a$ causes a greater fraction of the imposed thermal energy to be dissipated directly from the loop to the surrounding environment, thereby reducing the energy transported to and removed by the heat exchanger. Increasing $\lambda_p$, in contrast, reduces the effective thermal resistance of the powder-gap region and promotes heat removal through the heat exchanger. The two parameters therefore influence the partition of thermal energy between distributed environmental losses and heat-exchanger removal in opposite ways.

The similarity between the correlation patterns of $\overline{\Delta T}_{\mathrm{HX,term}}$ and $\overline{\dot{Q}}_{\mathrm{HX,term}}$ is also physically expected because the latter is calculated from the product of mass flow rate, heat capacity, and heat-exchanger temperature difference. Within the calibrated posterior region, the variability of the terminal-window mean heat-removal rate is therefore strongly linked to the variability of the terminal-window mean heat-exchanger temperature difference. Nevertheless, $\overline{\dot{Q}}_{\mathrm{HX,term}}$ retains a more direct engineering interpretation because it represents the mean rate at which thermal energy is removed through the heat exchanger during the terminal analysis window.

The hydraulic resistance coefficient exhibits positive associations with $\overline{\Delta T}_{\mathrm{HX,term}}$ and $\overline{\dot{Q}}_{\mathrm{HX,term}}$, with coefficients of $0.72$ and $0.45$, respectively. Although an increase in $C_k$ reduces the mass flow rate, it also increases the temperature change experienced by the coolant across the heat exchanger. Within the posterior-supported region, the increase in heat-exchanger temperature difference partially offsets the effect of the reduced mass flow rate, resulting in a moderate positive association with the derived terminal-window heat-removal rate. This behavior further illustrates that the posterior association coefficients represent the net response of the coupled system and should not be interpreted through a single parameter--response pathway in isolation.

The association map in Fig.~\ref{fig:ss} is consistent with the joint posterior distribution shown in Fig.~\ref{fig:post32-16}. In particular, the posterior samples of $h_a$ and $\lambda_p$ form a narrow, nearly one-dimensional ridge, indicating a pronounced compensation mechanism between distributed heat loss and heat transfer through the heat-exchanger powder region. Different combinations of $h_a$ and $\lambda_p$ can therefore produce similar measured temperature and flow responses even though the implied partition of thermal energy between the two heat-removal pathways is different.

The opposite associations of $h_a$ and $\lambda_p$ with the heat-exchanger indicators should consequently not be interpreted as two independent sensitivity effects. Instead, they largely reflect movement along the strongly coupled posterior ridge observed in Fig.~\ref{fig:post32-16}. The available transient observations constrain the combined heat-removal behavior of the loop more strongly than they independently identify the contributions of distributed ambient heat loss and powder-gap thermal conduction. This represents a practical parameter-identifiability limitation, or equifinality, rather than a failure of posterior convergence. The narrowness of the ridge indicates that the admissible combinations are strongly constrained, whereas its elongated structure indicates that the two parameters cannot be uniquely separated using the available response measurements.

The weak associations involving $P_b$ are also consistent with this coupled interpretation. Variations in the heater power bias can be partially compensated by changes in ambient heat loss, heat-exchanger conductance, and flow-dependent energy transport. Consequently, the measured transient responses may be reproduced accurately without uniquely identifying the heater power bias. This highlights an important distinction between predictive calibration and parameter identification: a calibrated model can provide reliable predictions even when individual physical parameters remain weakly identifiable or strongly correlated.

Overall, the posterior association analysis indicates that the NACIE system contains a hydraulically dominated response direction, most strongly associated with $C_k$ within the posterior-supported region, and a thermally coupled heat-removal direction governed by the compensatory interaction between $h_a$ and $\lambda_p$. The ability of the adaptive IUQ framework to reproduce both the strong hydraulic association and the narrow thermal compensation ridge demonstrates that the proposed refinement strategy preserves not only marginal parameter uncertainty, but also the system-level physical dependence structures embedded in the high-fidelity posterior.

\subsection{Computational efficiency analysis}

The efficiency of the proposed framework can be considered from two complementary perspectives: computational efficiency and statistical sampling efficiency. Computational efficiency concerns the cost of model evaluation, high-fidelity data generation, and surrogate-based likelihood evaluation, whereas statistical efficiency concerns how effectively posterior samples are generated for a given sampling effort. In the present application, however, the dominant computational burden arises from the high-fidelity Flownex SE simulations. Therefore, the quantitative comparison in this subsection focuses primarily on the reduction in the number of high-fidelity model evaluations.

One complete Flownex SE transient simulation requires approximately 120~s, whereas one evaluation of the trained PCA--GPR surrogate requires less than 0.01~s. The surrogate evaluation is therefore more than four orders of magnitude faster than the corresponding high-fidelity simulation. Consequently, the computational overhead associated with surrogate-based posterior sampling, candidate scoring, and surrogate prediction is small compared with the cost of generating the high-fidelity Flownex SE data. The simulation times reported in Table~\ref{tab:computational_efficiency_comparison} therefore account only for the dominant high-fidelity simulation cost.

\begin{table}[htbp]
\centering
\caption{Comparison of the high-fidelity Flownex SE simulation requirements for different IUQ strategies.}
\label{tab:computational_efficiency_comparison}
\resizebox{\textwidth}{!}{%
\begin{tabular}{
    l
    c
    S[table-format=3.1]
}
\toprule
{Method}
& {Number of Flownex SE simulations}
& {Estimated simulation time (CPU h)} \\
\midrule

Direct Flownex-based IUQ
& $10000$
& 333.3 \\

Fixed PCA--GPR surrogate IUQ~\cite{yao1}
& $1024$
& 34.1 \\

Proposed adaptive IUQ
& $288$
& 9.6 \\

\bottomrule
\end{tabular}
}
\end{table}

As shown in Table~\ref{tab:computational_efficiency_comparison}, direct Flownex-based IUQ requires approximately 10000 Flownex SE simulations, corresponding to approximately 333.3~h of high-fidelity simulation time. Although this approach provides the high-fidelity reference posterior used in the present assessment, its computational cost makes routine or repeated application difficult for complex transient thermal-hydraulic systems. A fixed PCA--GPR surrogate substantially reduces the online cost of posterior sampling by replacing repeated Flownex SE evaluations with inexpensive surrogate predictions. However, constructing a sufficiently accurate fixed surrogate may still require a large high-fidelity training dataset. For example, Yao et al.~\cite{yao1} used 1024 Flownex SE simulations to construct a high-accuracy PCA--GPR surrogate, corresponding to approximately 34.1~h of high-fidelity simulation time. Because these training samples are generated before the posterior-relevant region is identified, part of the computational effort may be allocated to regions that have limited influence on the final Bayesian inference. In contrast, the proposed adaptive IUQ framework requires only 288 Flownex SE simulations, corresponding to approximately 9.6~h of high-fidelity simulation time. Relative to direct Flownex-based IUQ, this represents a reduction of approximately 97.1\% in the number of high-fidelity model evaluations and an approximately 34.7-fold reduction in the dominant high-fidelity simulation cost. Relative to the fixed-surrogate strategy using 1024 training simulations, the proposed method reduces the high-fidelity data-generation requirement by approximately 71.9\%, corresponding to an approximately 3.6-fold improvement in high-fidelity data-generation efficiency. Importantly, this reduction is achieved while maintaining posterior accuracy comparable to the direct high-fidelity reference.

The statistical sampling efficiency of MCLMC was investigated in the preceding study by Yao et al.~\cite{yao1}, where its efficiency for surrogate-based posterior exploration was evaluated in comparison with alternative sampling strategies. Therefore, the computational comparison presented in Table~\ref{tab:computational_efficiency_comparison} does not repeat or include the sampler-level efficiency differences associated with MCLMC. Instead, the present analysis isolates the main computational bottleneck addressed by the proposed adaptive framework: the number of expensive high-fidelity Flownex SE simulations required to construct a surrogate with sufficient accuracy for posterior inference. This distinction is important because, once the PCA--GPR surrogate has been constructed, each surrogate evaluation requires less than 0.01~s, whereas a complete Flownex SE transient simulation requires approximately 120~s. Consequently, although the statistical efficiency of the posterior sampler remains important, its computational contribution is secondary to the cumulative cost of high-fidelity data generation in the present application. The efficiency gains reported in Table~\ref{tab:computational_efficiency_comparison} therefore refer specifically to the reduction in high-fidelity model evaluations and the corresponding Flownex SE simulation time.

Overall, the proposed framework improves efficiency through two complementary mechanisms: statistically efficient posterior exploration using MCLMC and computationally efficient high-fidelity data generation through posterior-guided adaptive refinement. For the present NACIE application, the latter provides the dominant practical benefit by substantially reducing the number of expensive Flownex SE simulations required to achieve reliable posterior inference.







\subsection{Posterior Predictive Assessment of Transient Responses}
To further assess whether the adaptively refined surrogate can reproduce the calibrated transient responses, posterior predictive validation is performed using posterior samples obtained from the adaptive IUQ framework. Unlike prior-based forward uncertainty propagation, which reflects uncertainty over the full admissible parameter space, the present analysis focuses on the posterior-relevant region identified after assimilating the NACIE measurements. This is consistent with the objective of the proposed adaptive framework, which is to improve surrogate fidelity in regions that are most influential for Bayesian calibration. For each selected posterior sample, the transient responses predicted by the adaptive PCA-GPR surrogate are compared with the corresponding Flownex SE simulations. The comparison is performed for all output channels, including the temperature responses and loop mass flow rate. The resulting posterior predictive intervals are used to evaluate whether the adaptive surrogate preserves the calibrated output uncertainty and reproduces the high-fidelity transient behavior in the posterior region.

\subsubsection{Calibration-channel comparison}

\begin{figure}[t!]
    \centering
    \includegraphics[width=1\linewidth]{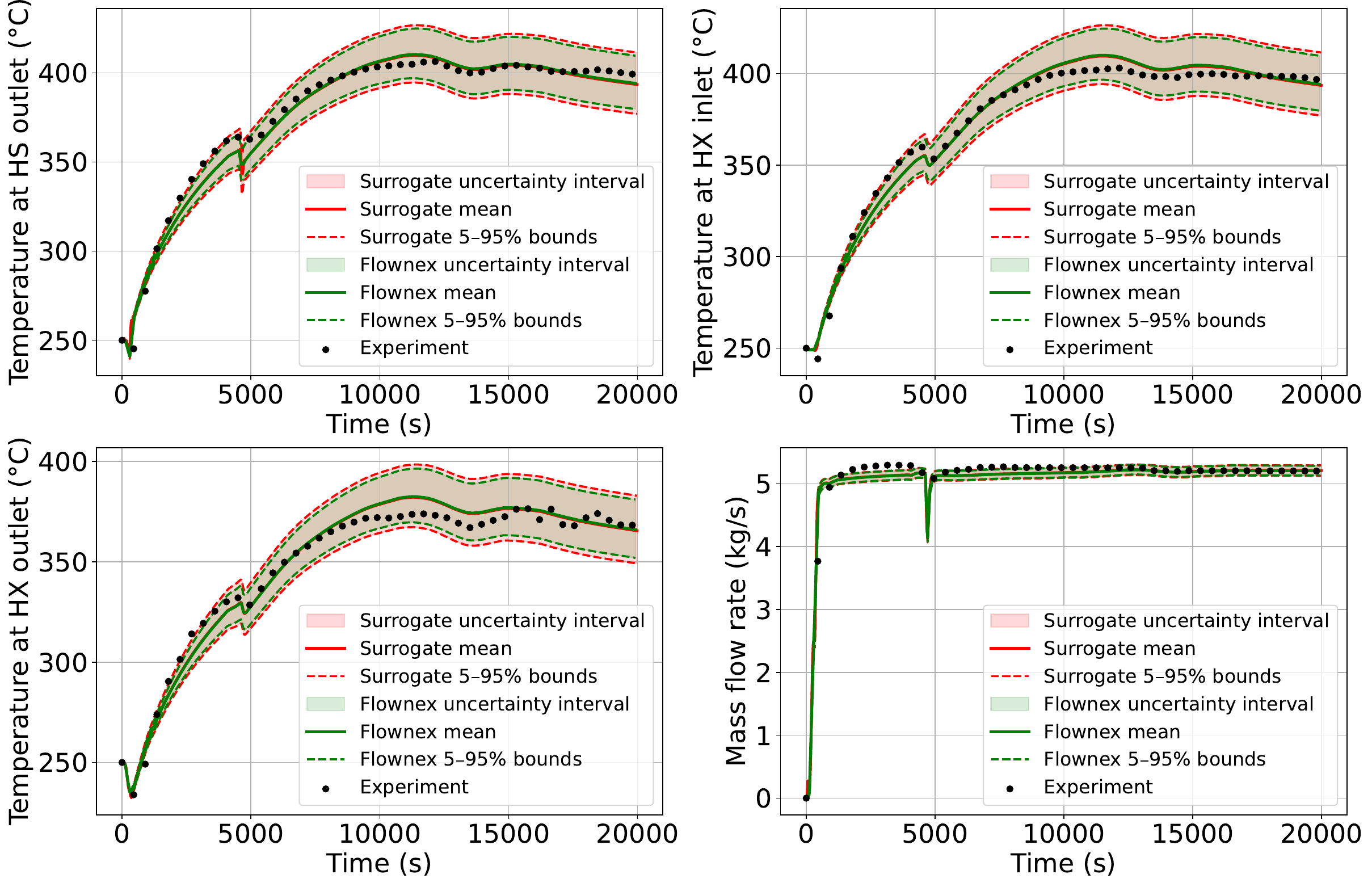}
    \caption{\centering Posterior predictive comparison for the calibration channels. The adaptive PCA-GPR surrogate predictions are compared with direct Flownex SE simulations using posterior samples obtained from the adaptive IUQ framework. The shaded regions denote the 5--95\% posterior predictive intervals, and the black points denote experimental observations.}
    \label{fig:iuq}
\end{figure}

Figure~\ref{fig:iuq} compares the posterior predictive responses obtained from the adaptive PCA-GPR surrogate and direct Flownex SE simulations for the four calibration channels used in the likelihood function. Overall, the adaptive surrogate shows very good agreement with the direct Flownex SE results for both the posterior mean responses and the uncertainty intervals. For the three temperature channels, including the HS outlet, HX inlet, and HX outlet temperatures, the surrogate posterior mean closely follows the corresponding Flownex mean throughout the transient. The initial heating stage, the gradual increase in temperature, and the later transient variations are all reproduced consistently by the surrogate model.

A noticeable feature is that the posterior predictive intervals obtained from the adaptive surrogate are slightly wider than those from the direct Flownex SE simulations. This behavior is expected because the surrogate prediction includes an additional epistemic uncertainty contribution from the PCA-GPR emulator. More importantly, the surrogate-based 5--95\% intervals almost completely cover the corresponding Flownex-based intervals over the full transient period. This indicates that the adaptive surrogate does not underestimate the output uncertainty in the posterior-relevant region. Instead, it provides a slightly more conservative but still highly consistent representation of the high-fidelity posterior predictive uncertainty.

The agreement is also clear for the loop mass-flow response. The adaptive surrogate accurately captures the rapid initial increase in mass flow and the subsequent nearly steady natural-circulation behavior. Although small local deviations can be observed near sharp transient features, the surrogate mean remains close to the Flownex mean, and the surrogate predictive interval continues to enclose the Flownex uncertainty band for most of the simulation period. This is important because the mass-flow response is strongly coupled with the thermal evolution of the loop and is sensitive to hydraulic resistance, buoyancy feedback, and heat-removal behavior. The close agreement therefore suggests that the adaptively refined surrogate preserves the dominant coupled thermal--hydraulic behavior of the high-fidelity Flownex model in the posterior region.

The experimental observations are generally located within or close to the posterior predictive intervals for all calibration channels. Since these four responses are directly used in the likelihood function, the purpose of this comparison is not to provide an independent validation, but to assess whether the adaptive surrogate can reproduce the high-fidelity calibrated responses after Bayesian inference. The close overlap between the surrogate-based and Flownex-based posterior predictive bands demonstrates that the posterior-guided adaptive refinement successfully improves surrogate fidelity in the regions most relevant to the posterior distribution. Consequently, the adaptive PCA-GPR surrogate can replace repeated Flownex SE evaluations during posterior prediction while introducing negligible bias and maintaining a conservative estimate of output uncertainty.

\subsubsection{Hold-out validation on the HS inlet temperature}

\begin{figure}[t!]
    \centering
    \includegraphics[width=1\linewidth]{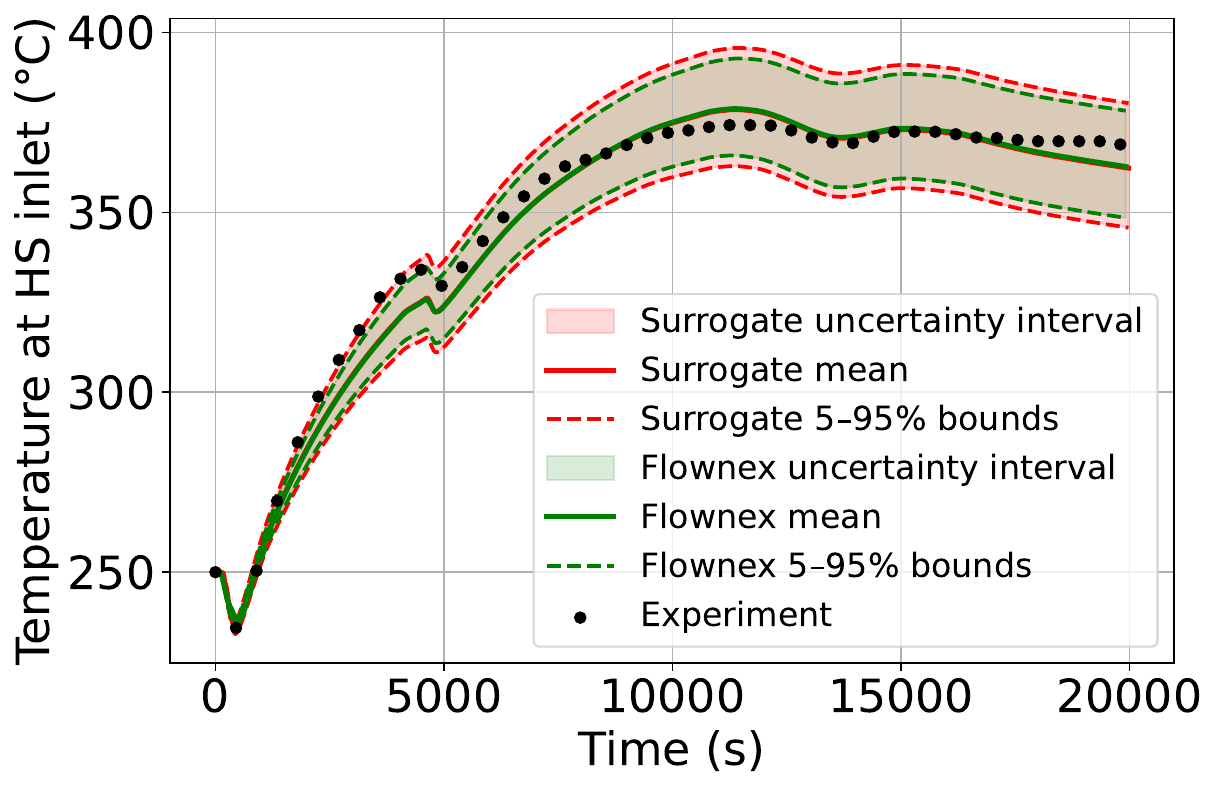}
    \caption{\centering Posterior predictive cross-validation of the adaptive surrogate for the hold-out HS inlet temperature response. }
    \label{fig:iuq-vv}
\end{figure}

Figure~\ref{fig:iuq-vv} presents the hold-out posterior predictive validation using the HS inlet temperature, which is not included in the likelihood function during Bayesian calibration. Therefore, this response provides an independent check of the predictive capability of the adaptive IUQ framework for an unassimilated transient channel. Overall, the adaptive PCA--GPR surrogate shows good agreement with the direct Flownex SE simulations for the HS inlet temperature. The surrogate posterior mean closely follows the Flownex mean over the full transient, including the initial temperature drop, the subsequent heating stage, and the later-stage transient evolution. Similar to the calibration-channel results, the surrogate predictive interval is slightly wider than the Flownex interval, mainly because the surrogate includes additional emulator uncertainty. Nevertheless, the surrogate-based 5--95\% interval almost fully covers the corresponding Flownex uncertainty band. The experimental observations are also generally located within or close to the posterior predictive interval. Since this channel is not used for likelihood evaluation, the agreement indicates that the calibrated posterior and the adaptive surrogate are not merely reproducing the assimilated outputs, but also retain predictive consistency for an unused thermal response. This hold-out validation further supports the applicability of the posterior-guided adaptive IUQ framework for transient multi-output thermal--hydraulic prediction.

\section{Conclusions}

This study developed a posterior-guided adaptive surrogate refinement framework for data-efficient Bayesian calibration of computationally expensive transient thermal--hydraulic models. The central objective was not to construct a uniformly accurate surrogate over the complete prior domain, but to preserve the posterior distribution and calibrated transient responses under a limited high-fidelity simulation budget. To achieve this objective, the framework combines PCA-based compression of multi-channel transient outputs, GPR surrogate modeling, MCLMC posterior sampling, a persistent extensible Sobol candidate reservoir, posterior-informed local--global classification, and uncertainty--novelty--diversity-based batch selection.

The NACIE application demonstrates that posterior fidelity can be achieved without uniformly resolving the entire admissible parameter space. Using 288 Flownex SE evaluations, the adaptive surrogate reproduced the principal characteristics of the direct simulator-based posterior, including parameter means, uncertainty levels, percentile ranges, marginal distributions, and strongly correlated joint structures. The posterior predictive responses for the calibration channels were also closely reproduced, while the HS inlet temperature provided an out-of-likelihood assessment showing that the calibrated parameter distribution retained predictive consistency for an unassimilated response channel. These results indicate that posterior-guided refinement can concentrate model evaluations in regions where surrogate errors are most consequential for Bayesian inference.

The system-level association analysis further showed that preservation of the joint posterior has direct physical significance. Within the experimentally constrained posterior region, the terminal-window natural-circulation mass flow rate was most strongly associated with the hydraulic resistance coefficient $C_k$. In contrast, the heat-exchanger response was characterized by a narrow compensation direction involving the ambient heat-transfer coefficient $h_a$ and the powder-gap thermal conductivity $\lambda_p$. This structure indicates that the measurements strongly constrain the combined heat-removal behavior of the loop while providing weaker information for independently separating distributed ambient losses from heat removal through the heat exchanger. Consequently, reliable predictive calibration can be achieved even when individual physical parameters remain strongly correlated or only partially identifiable.

Relative to a fixed PCA--GPR surrogate constructed using 1024 simulations, the adaptive framework reduced the required high-fidelity training evaluations by 71.9\% while retaining comparable posterior accuracy. Relative to a 10000-evaluation direct-simulation reference, the method reduced the number of Flownex SE evaluations by approximately 97.1\%, corresponding to an approximately 34.7-fold reduction in reference-model evaluation cost. These values quantify the reduction in high-fidelity data-generation cost and should be distinguished from total workflow wall-clock acceleration, which also depends on surrogate retraining, posterior sampling, and adaptive-selection overhead.

More broadly, the results demonstrate that global surrogate accuracy is not necessarily the most relevant objective for surrogate-assisted IUQ. A surrogate may be unnecessarily accurate in regions with negligible posterior probability while remaining insufficiently accurate in the limited region that controls the likelihood. By directly linking refinement to posterior geometry and predictive uncertainty, the proposed framework provides a practical strategy for allocating expensive simulations according to their value for probabilistic inference rather than their contribution to uniform approximation.

The present conclusions are limited to a single system-scale transient benchmark and a working likelihood that mitigates, but does not explicitly model, temporal and cross-channel residual dependence. The hold-out analysis evaluates an unassimilated response from the same transient rather than an independent operating condition, and the scalar residual-error factor does not represent a structured model-discrepancy process. Future work will therefore investigate multiple-transient and hierarchical calibration, structured residual models, multi-modal posterior refinement, and applications involving coupled system-code/CFD simulations. Within these limitations, the proposed framework establishes a data-efficient basis for repeated uncertainty-aware model updating and for subsequent reliability, safety-margin, and digital-assessment applications involving expensive transient system models.


\section*{CRediT authorship contribution statement}
Zikang Yao: Writing – original draft, Formal analysis, Methodology, Investigation, Visualization, Data curation.  
Eugene Yee: Supervision, Conceptualization, Writing – review \& editing, Resources, Validation.  
Fue-Sang Lien: Funding acquisition, Project administration, Supervision, Writing – review \& editing.

\section*{Declaration of competing interest}
The authors declare that they have no known competing financial interests or personal relationships that could have appeared to influence the work reported in this paper.

\section*{Acknowledgments}
We acknowledge the support of the Natural Sciences and Engineering Research Council of Canada (NSERC) and Canadian Nuclear Safety Commission (CNSC), [funding reference number ALLRP 580447-2022].

\section*{Data availability}
The authors do not have permission to share data.

\bibliographystyle{elsarticle-num}   

\bibliography{references2} 

\end{document}